\documentclass{article}

\usepackage{microtype}
\usepackage{graphicx,tabularx}
\usepackage{subcaption}
\usepackage{booktabs,amsthm,epstopdf} 
\usepackage{multirow, balance}
\usepackage{hyperref}

\usepackage{amsmath}
\theoremstyle{definition}
\newtheorem{defn}{Definition}[]

\usepackage[accepted]{icml2019}

\icmltitlerunning{DNN Digital Passport for ICML 2019}

\begin{document}
\newcolumntype{C}[1]{>{\centering\arraybackslash}p{#1}}

\twocolumn[
\icmltitle{Digital Passport: A Novel Technological Strategy \\ for  Intellectual Property Protection of Convolutional Neural Networks}


\begin{icmlauthorlist}
\icmlauthor{Lixin Fan}{}
\icmlauthor{Kam Woh Ng}{UnivM}
\icmlauthor{Chee Seng Chan}{UnivM}
\end{icmlauthorlist}

\icmlaffiliation{UnivM}{Center of Image and Signal Processing, Faculty of Computer Science and Information Technology, University of Malaya, 50603 Kuala Lumpur, Malaysia}

\icmlcorrespondingauthor{Lixin Fan}{lixin.fan01@gmail.com}
\icmlcorrespondingauthor{Chee Seng Chan}{cs.chan@um.edu.my}

\icmlkeywords{Machine Learning, ICML}

\vskip 0.3in
]



\printAffiliationsAndNotice{\icmlEqualContribution} 

\begin{abstract}
In order to prevent deep neural networks from being infringed by unauthorized parties, we propose a generic solution which embeds a designated digital passport into a network, and subsequently, either \textit{paralyzes} the network functionalities for unauthorized usages or \textit{maintain} its functionalities in the presence of a verified passport. Such a desired network behavior is successfully demonstrated in a number of implementation schemes, which provide reliable, preventive and timely protections against tens of thousands of fake-passport deceptions. Extensive experiments also show that the deep neural network performance under unauthorized usages deteriorate significantly (e.g. with 33\% to 82\% reductions of CIFAR10 classification accuracies), while networks endorsed with valid passports remain intact. 
\end{abstract}

\section{Introduction}
\label{sect:intro}

While Machine Learning as a Service (MLaaS) has emerged as a viable and lucrative business model, there is an urgent need to protect deep neural networks (DNN) from being used, copied and re-distributed by unauthorized parties (i.e. intellectual property infringement). Recently, for instance, digital \textit{watermarking} techniques have been adopted to embed watermarks into DNN models during the training stage. Subsequently, ownerships of these models are verified by the detection of the embedded watermarks, which are supposed to be robust to multiple types of DNN modifications such as fine-tuning, pruning and watermark overwriting \cite{EmbedWMDNN_2017arXiv,AdStitch_2017arXiv,TurnWeakStrength_Adi2018arXiv}.  

\begin{figure}[t]
	\centering
	{\includegraphics[keepaspectratio=true, scale = 0.9]{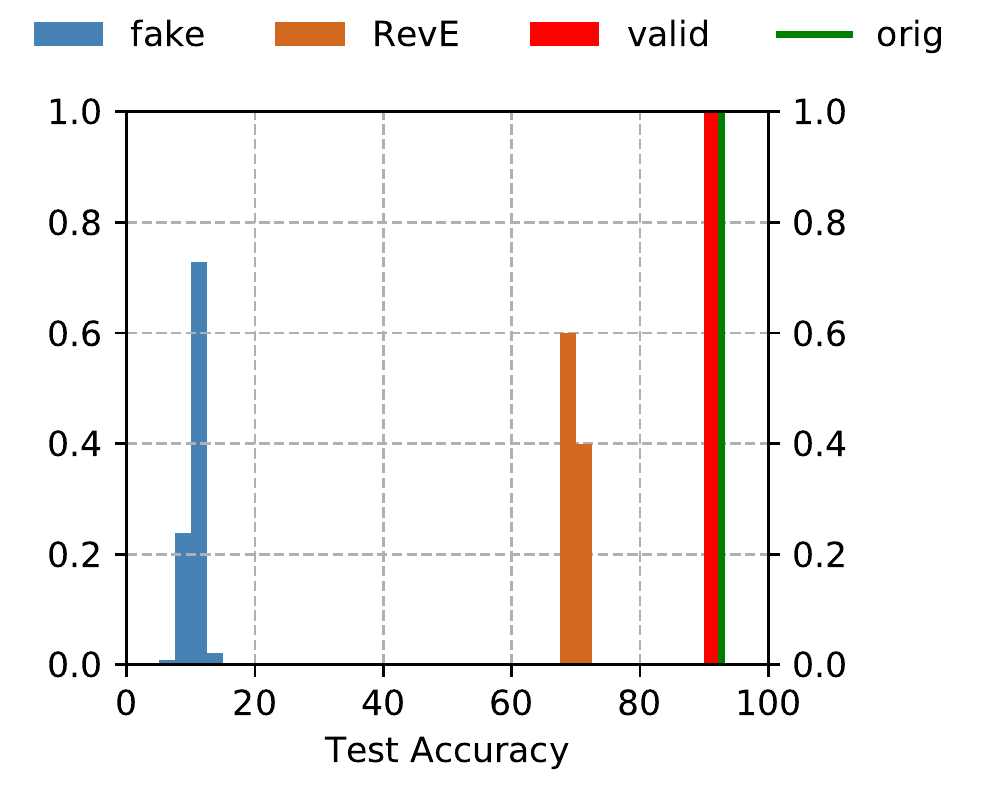}}\vspace{-6pt}
	\caption{A comparison of CIFAR10 classification test accuracies of original network (``orig'') against protected network with valid passports (``valid''), protected network with fake passports  (``fake''), and network with hidden parameters reverse-engineered (``RevE''). It shows that the proposed method is resilient against various attacks and provides a proactive IPR protection in opposition to infringements by paralyzing the network performance (e.g. from 92\% $\to$ $\pm$ 15\%).}
\label{fg:teaser}
\end{figure}

The principle of digital watermarking for neural network models is mainly inspired by the protection of intellectual property right (IPR) of digital media such as images or videos, which are being processed by media processors e.g. photo galleries or video players. Nevertheless, this approach disregards a unique and fundamental feature of neural network models ---   themselves are information processors that fulfill certain tasks, including the classification, detection or manipulation of input information.  
Therefore, a novel and preferable strategy in neural network IP protection is \textit{to paralyze the DNN models against unauthorized parties}, while maintaining their normal functionalities for lawful usage. To this end, a designated digital entity must be presented to endorse authorized usage of the protected network in question.  We refer to this type of protection entities as \textit{digital passports} and, correspondingly, the process of  embedding digital passports into a DNN model is referred to as digital \textit{passporting}. Within this paper, we shall illustrate how to embed digital passports so that the resulting DNN models are both \textit{functionality-preserving} and \textit{well-protected} (see definitions in Section \ref{sect:formulation}).  

On the one hand, digital passporting bears a similarity to digital watermarking --- they both embed certain digital entities into DNN models during training sessions. In terms of the IPRs protection, however, embedded watermarks only enable the verification of the ownership of DNN models. One has to rely on the government investigation and enforcing actions to discourage the infringement of IPRs. More often than not, this kind of juridical protection might be unreliable, costly and long-overdue.   On the other hand, passports-protected DNN models will not function normally unless valid passports are provided, thus immediately preventing the unlawful usages of the networks 
with no extra costs. Indeed, we regard this proactive protection the most prominent advantage of digital passporting over digital watermarking. For instance, as shown in Figure \ref{fg:teaser} of CIFAR10 classification performances, the protected networks with valid passports demonstrated almost identical accuracies as that of the original network, while the same networks presented with fake passports merely achieved about 10\% classification rates. Moreover, even if a computationally demanding reverse-engineering algorithm was used to recover protected network parameters (see Section \ref{subsect:reverse}), the best accuracies obtained was no more than 70\%, substantially inferior to that of the protected network i.e. 92\%. 

To our best knowledge, the present paper is the first work that shows how to embed and use digital passports to prevent DNN models from being infringed by unauthorized parties. Section \ref{related_work} reviews recent digital watermarking techniques developed for DNNs. Section \ref{sect:digPass} details the DNN architectures for embedding and verifying digital passports of target DNN models. Section \ref{sect:pass-as-watermark} explains the digital passports as watermarks; with extensive experiment results in Section \ref{sect:exper} demonstrates the efficacy of the passporting method. 

In summary, the contributions of this paper are as follows: 
\begin{itemize}
\item We renovate the paradigm of digital watermarking based neural network IP protection, by proposing a digital passporting based strategy which provides \textit{reliable, preventive} and \textit{timely} IP protection at virtually \textit{no extra cost} for all neural networks. 

\item This paper formulates the digital passporting problem and proposes a generic solution as well as concrete implementation schemes that embed digital passports into DNN models through dedicated passporting layers (Section \ref{sect:digPass}; Figure \ref{fig:nn-pass-arch}). The embedded passports prevent the unauthorized network usage (infringement) by paralyzing the networks while maintaining its functionality for verified users (Section \ref{sect:expt-pass}; Figure \ref{fig:hist-resnet-cifar10}).  

\item Our work shows that the embedded passports also verify the ownership of networks, in case that the DNN hidden parameters are illegally disclosed or reverse-engineered while public parameters are plagiarized (Section \ref{sect:pass-as-watermark}). 

\end{itemize}

\subsection{Related work}\label{related_work}

\cite{EmbedWMDNN_2017arXiv} was probably the first work that proposed a general framework to embed watermarks into DNN models by imposing an additional regularization term on the weights parameters i.e. $E_R(w)$, which is dependent on the watermarks to be embedded.
It was shown that the performances of the original networks (for image classifications) were not affected by the embedded watermarks, and the ownership of network models were robustly verified against a variety of modifications like fine-tuning and pruning. 

However, the aforementioned method was limited in the sense that one has to access all the network weights in question to extract the embedded watermarks (i.e. white-box setting). Therefore, \cite{AdStitch_2017arXiv}  proposed to embed watermarks in the classification labels of adversarial examples, so that the watermarks can be extracted remotely through a service API without the need to access the network internal weights parameters (i.e. black-box setting). Later, \cite{TurnWeakStrength_Adi2018arXiv} proved that embedding watermarks in the networks' (classification) outputs is actually a designed \textit{backdooring} and provided theoretical analysis of performances under various conditions. Also in black-box and white box settings, \cite{DeepSigns_2018arXiv,DeepMarks_2018arXiv,8587745} demonstrated how to embed  watermarks (or fingerprints), that are robust to watermark overwriting, model fine-tuning and pruning. Noticeably, a wide variety of DNN architectures such as Wide Residual Networks (WRNs) and Convolutional Neural Networks (CNNs) were investigated.  

\cite{mun2017robust} employed two CNN networks to embed a one-bit watermark in a single image block; while \cite{vedranwifs} investigates a new family of transformation based on Deep Learning networks for blind image watermarking. In one of the latest work, IBM team -  \cite{ProtectIPDNN_Zhang2018} proposed to use three types of watermarks (i.e. \textit{content, unrelated} and \textit{noise})  and demonstrated their performances with MNIST and CIFAR10 datasets. Lastly, \cite{zhu2018hidden} proposed an end-to-end trainable framework, HiDDeN for data hiding in color images based on CNNs and Generative Adversarial Network and may be applied to watermarking and steganography.


\section{Digital Passport for Deep Neural Networks}
\label{sect:digPass}

The ultimate goal of digital passporting is to \textit{design} and \textit{train} DNN models in a way such that, only if a designated digital passport (or signature) is presented, will the protected networks function normally. Otherwise, the functionalities of the original networks will be paralyzed.  In the following, we shall first formulate the desired characteristics of the DNN protected by the digital passports, and illustrate a generic solution as well as concrete implementation schemes that are employed to embed the digital passports into a convolutional neural network (CNN)\footnote{This paper shall only focus on digital passporting of CNNs.  The passporting of other network architectures such as GANs is out of the scope of this paper, and will be reported elsewhere.}.

\subsection{Problem formulation}\label{sect:formulation}

Let $\mathcal{N}$ denote a CNN model to be protected by a \textit{secret} digital passport $p$, after a training or passporting process, the network embedded with the passport is denoted by $\mathcal{N}[p]$. The inference of such a protected model can be characterized as a process $M$ that \textit{modifies} network behavior according to the running-time digital passport $s$: 
\begin{align}\label{eq:dpp-formula}
M( \mathcal{N}[p], s) = \left\{ \begin{array}{cc}
\mathcal{M}_p, & \text{if }s=p,  \\
\mathcal{M}_{\bar{p}}, & \text{otherwise},  \\
\end{array} \right.
\end{align}
in which $\mathcal{M}_p$ is the network performance with passport correctly verified, and $\mathcal{M}_{\bar{p}}$ is the performance with the incorrect passports i.e. $\bar{p} \neq p$. 

The properties  of  $M( \mathcal{N}[p], s)$ defined below are desired for the sake of intellectual property protection: 
\begin{defn}
	If $s=p$, the performance $ \mathcal{M}_p $ should be as close as possible to that of the original network $\mathcal{N}$. Specifically, if the performance \textit{inconsistency} between $ \mathcal{M}_p $ and that of $\mathcal{N}$ is smaller than a desired threshold e.g. $\tau_d=1\%$, then the protected network is called \textit{functionality-preserving}.
\end{defn}

\begin{defn}
If $s\neq p$, on the other hand, the performance $\mathcal{M}_{\bar{p}}$  should be as far as possible to that of $\mathcal{M}_{{p}}$. The discrepancy between $\mathcal{M}_{{p}}$ and  $\mathcal{M}_{\bar{p}}$ therefore can be defined as the \textit{protection-strength}. Moreover, if the \textit{strength} is larger than a desired threshold e.g. $\tau_s=50\%$, then the network in question is called \textit{well-protected}. 
\end{defn}

\subsection{A generic solution}

In order to modify the behavior of the CNN as formulated in (\ref{eq:dpp-formula}), we propose the following generic solution: we first \textit{partition} the set of CNN parameters into two non-overlapping subsets i.e. $W = \{W_p, W_h\}$, and use the \textit{public parameters} $W_p$, together with the secret passport $p$, to determine the \textit{hidden\footnotemark parameters} $W_h$.\footnotetext{In this work, traditional hidden layer parameters are considered as public parameters unless they are protected by (\ref{eq:hidden-from-public-key}).} Formally, this principle 
can be illustrated as follows: 
\begin{align}\label{eq:hidden-from-public-key}
W_h = \mathcal{F} ( W_p, p ),
\end{align}
in which $\mathcal{F}$ denotes a set of mathematical functions that calculate the values of $W_h$ from the given $W_p$ and $p$.  We refer to $\mathcal{F}$ as \textit{passport functions} in the rest of the present paper. 

The \textit{learning} or the \textit{digital passporting} of a CNN model therefore involves the optimization of  the network objective function e.g. to minimize the \textit{cross entropy} loss by adjusting public parameters $W_p$. The hidden parameters $W_h$ are no longer trainable, instead, they are directly updated according to (\ref{eq:hidden-from-public-key}).


During the \textit{inferencing} stage, the hidden parameters is computed with the running-time passport i.e. $W_h = \mathcal{F} ( W_p, s )$. Clearly, only if the designated secret passport is provided $s=p$ , will the hidden parameters  $W_h$ be set correctly. Otherwise, the CNN functionalities will be paralyzed due to incorrect values of hidden parameters $W_h$. 

\begin{table}[t]
	\centering
	
	\begin{tabular}{|c|c|c|}
		\hline
		& $\mathcal{F}_{\gamma}(W^l_p, p^l_{\gamma})$ & $\mathcal{F}_{\beta}(W^l_p, p^l_{\beta})$  \\		
		\hline
		\hline
		V1 & $\text{Avg}( W^l_p * p^l_{\gamma} )$  &  B  \\
		\hline
		V2 &  B &  $\text{Avg}( W^l_p * p^l_{\beta} )$  \\
		\hline
		V3 &  $\text{Avg}( W^l_p * p^l_{\gamma} )$  &  $\text{Avg}( W^l_p * p^l_{\beta} )$ \\
		\hline
	\end{tabular}
	\caption{Three different choices of passport functions (B means the parameter is a trainable variable as in standard Batch Normalization layer, $*$ denotes the convolution of $p$ with kernel $W$, and $\text{Avg}$ denotes the channel-wise average of convolution outputs). }
	\label{tab_pass_func}
\end{table} 

\subsection{Concrete implementations}

A concrete implementation of the generic solution therefore has to answer two questions: 
\begin{enumerate}
	\item How to partition the CNN parameters into 
	$W_p, W_h$? 	
	\item Which mathematical functions $\mathcal{F}$ are to be used? 
\end{enumerate}
We shall illustrate below a number of implementation schemes, which have their respective answers to the above questions. 
In particular, the implementations are inspired by the commonly adopted batch normalization technique, which essentially applies the channel-wise  linear transformation to the inputs \cite{BatchNorm_IoffeS15}. 

\begin{figure}[t]
	\centering
	\includegraphics[width=.35\textwidth,height=0.39\textwidth]{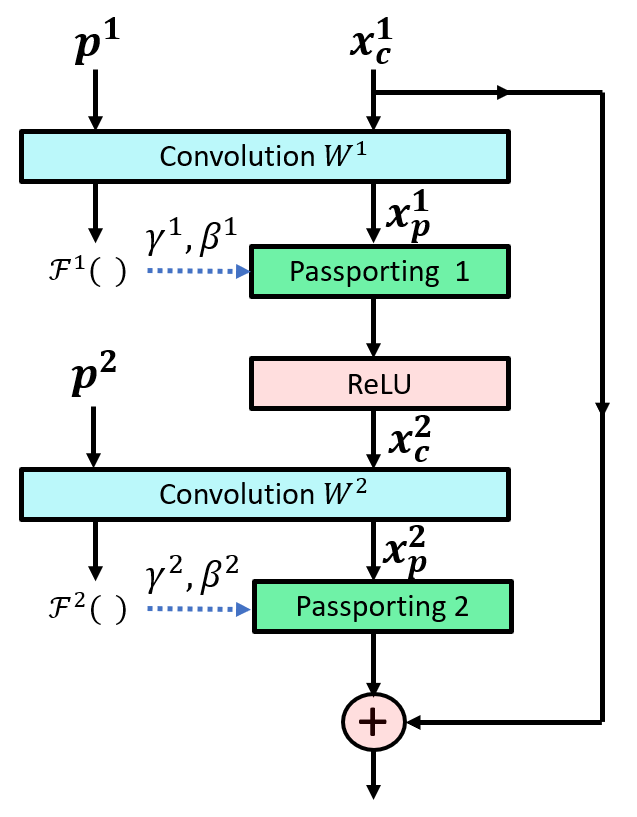}
	\caption{An example \textit{ResNet} layer that consists of two convolution layers, two digital passporting layers and the ReLU activation. 
		$x_c^l$ and $x_p^l$ are, respectively, inputs to the convolution and passporting layers, $p^l = \{p^l_{\gamma}, p^l_{\beta}\}$ is the \textit{digital passports} of corresponding layers.
		$\mathcal{F}$ is a passport function to compute the hidden parameters (i.e. $\gamma$ and $\beta$) of the passporting layers. Note that secret passports $p^l$ are used during the \textit{learning} stage, while running-time passports $s$ are used during the \textit{inferencing} stage. }
\label{fig:nn-pass-arch}
\end{figure}

In this work, we propose to append after each convolution or fully connected layer  a \textit{digital passporting layer},  whose scale factor $\gamma$ and bias shift term $\beta$ are dependent on both the weights of the preceding layer $W_p$ and the secret passport $p$ as follows: 
\begin{align}
P^l( x_p ) = \gamma^l x_p + \beta^l, \label{eq:plinear}\\
\gamma^l = \mathcal{F}_{\gamma}(W^l_p, p^l_{\gamma}), \label{eq:pass-gamma} \\ 
\beta^l = \mathcal{F}_{\beta}(W^l_p, p^l_{\beta}),\label{eq:pass-beta}
\end{align}
in which $l$ denotes the layer number, $x_p$ is the input to the passport layer, $P()$ is the corresponding linear transformation outputs, $\mathcal{F}_{\gamma}$ and $\mathcal{F}_{\beta}$ are the passport functions, while $p^l_{\gamma}$ and $p^l_{\beta}$ are the passports used to derive scale factor and bias term respectively. 
Figure \ref{fig:nn-pass-arch} depicts the architecture of digital passport layers used in a \textit{ResNet} layer and Table \ref{tab_pass_func} summarizes different choices of passport functions $\mathcal{F}_{\gamma}, \mathcal{F}_{\beta}$ that have been employed in our work. 

The practical choice of formula (\ref{eq:plinear}),(\ref{eq:pass-gamma}) and (\ref{eq:pass-beta}) is inspired by the Batch Normalization (BN) layer  \cite{BatchNorm_IoffeS15}, and that is also why V1,V2 still train $\gamma$ or $\beta$ following BN. 
Respective performances of these three choices, against different attacks, are illustrated and discussed in the following sections (see Section \ref{subsect:reverse} and \ref{sect:exper}). 

	\begin{figure*}[ht]
		\begin{subfigure}{\textwidth}
			\centering
			{\includegraphics[keepaspectratio=true, scale = 0.25]{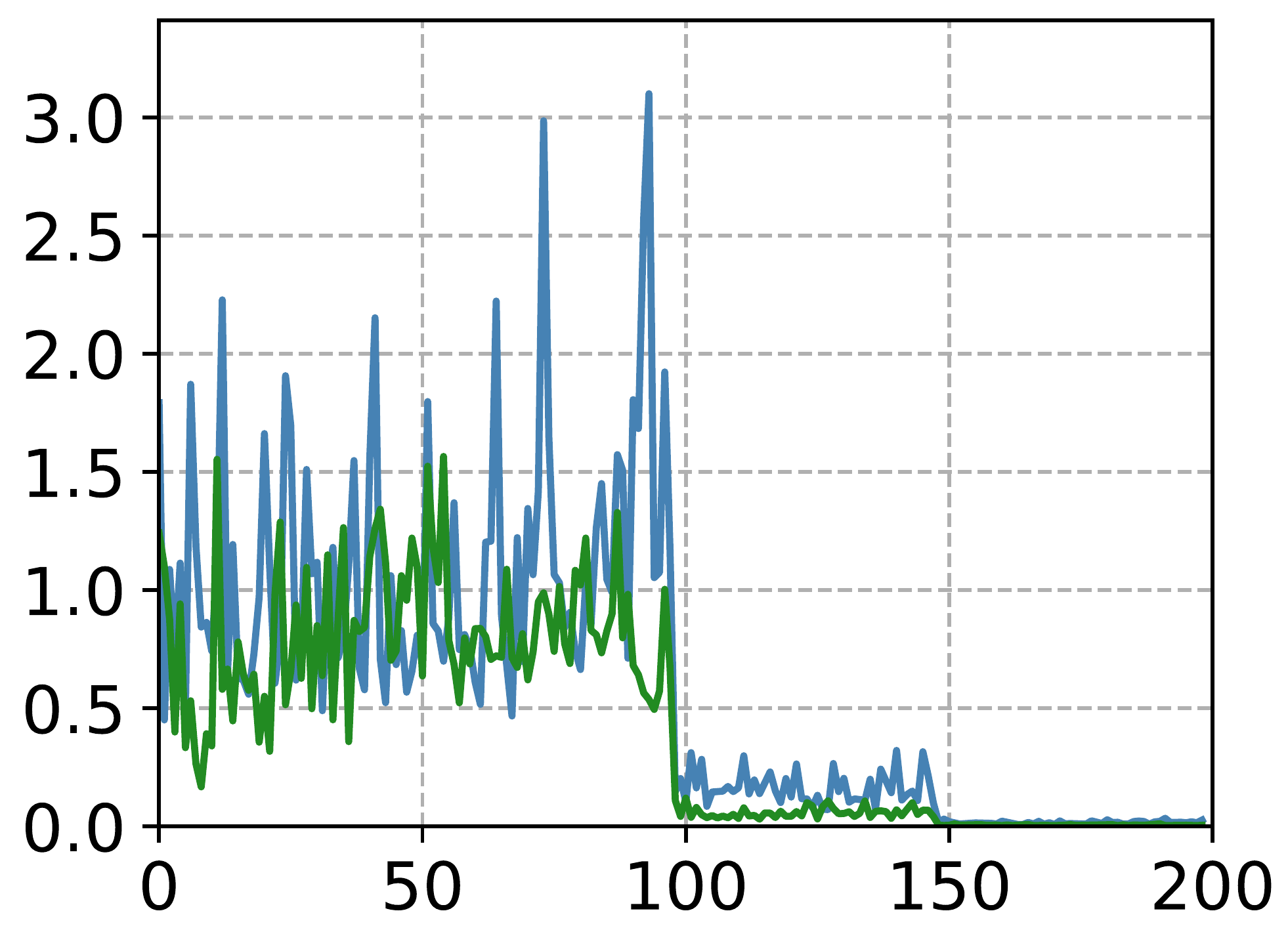}}
			{\includegraphics[keepaspectratio=true, scale = 0.25]{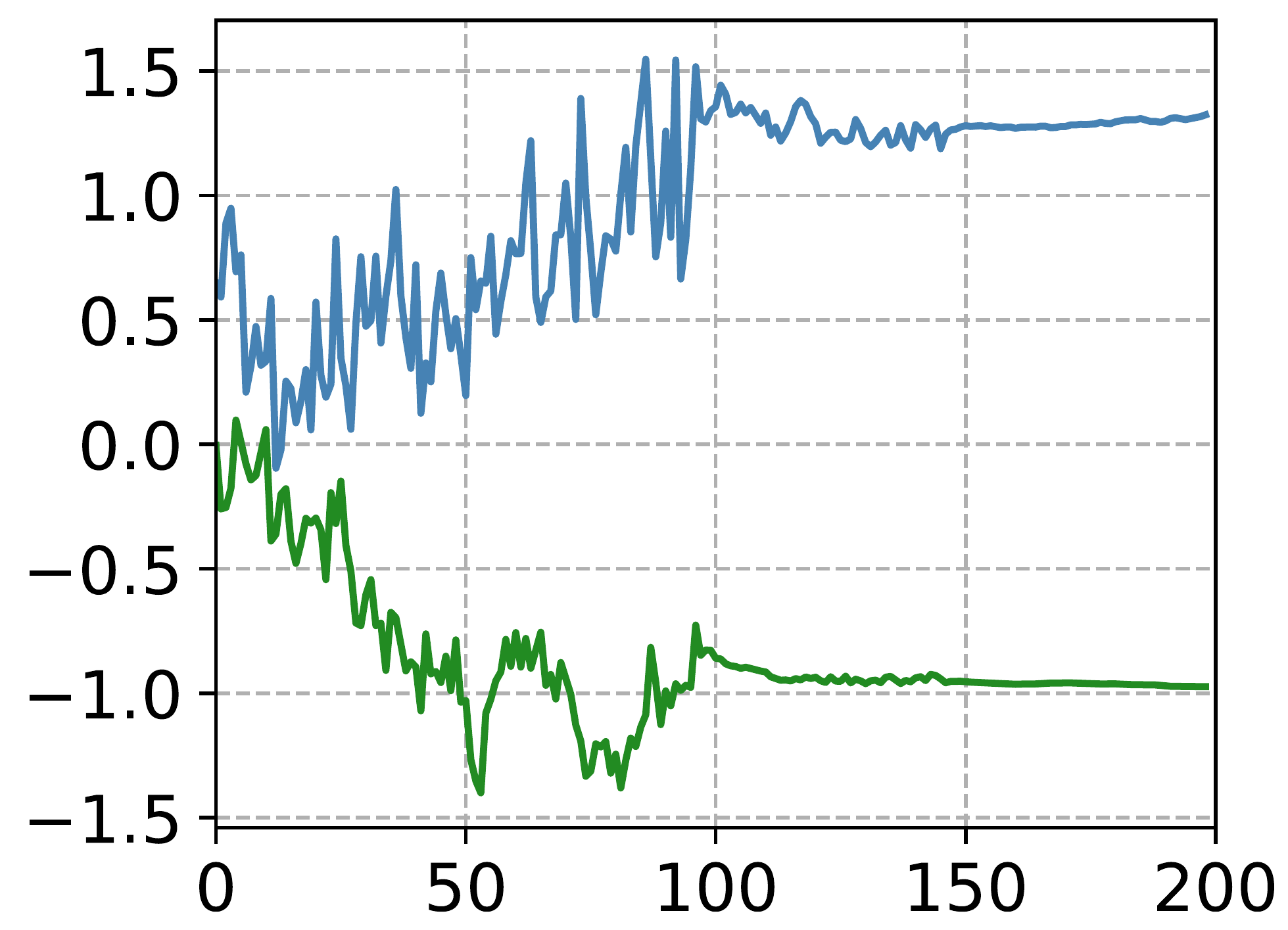}}
			{\includegraphics[keepaspectratio=true, scale = 0.25]{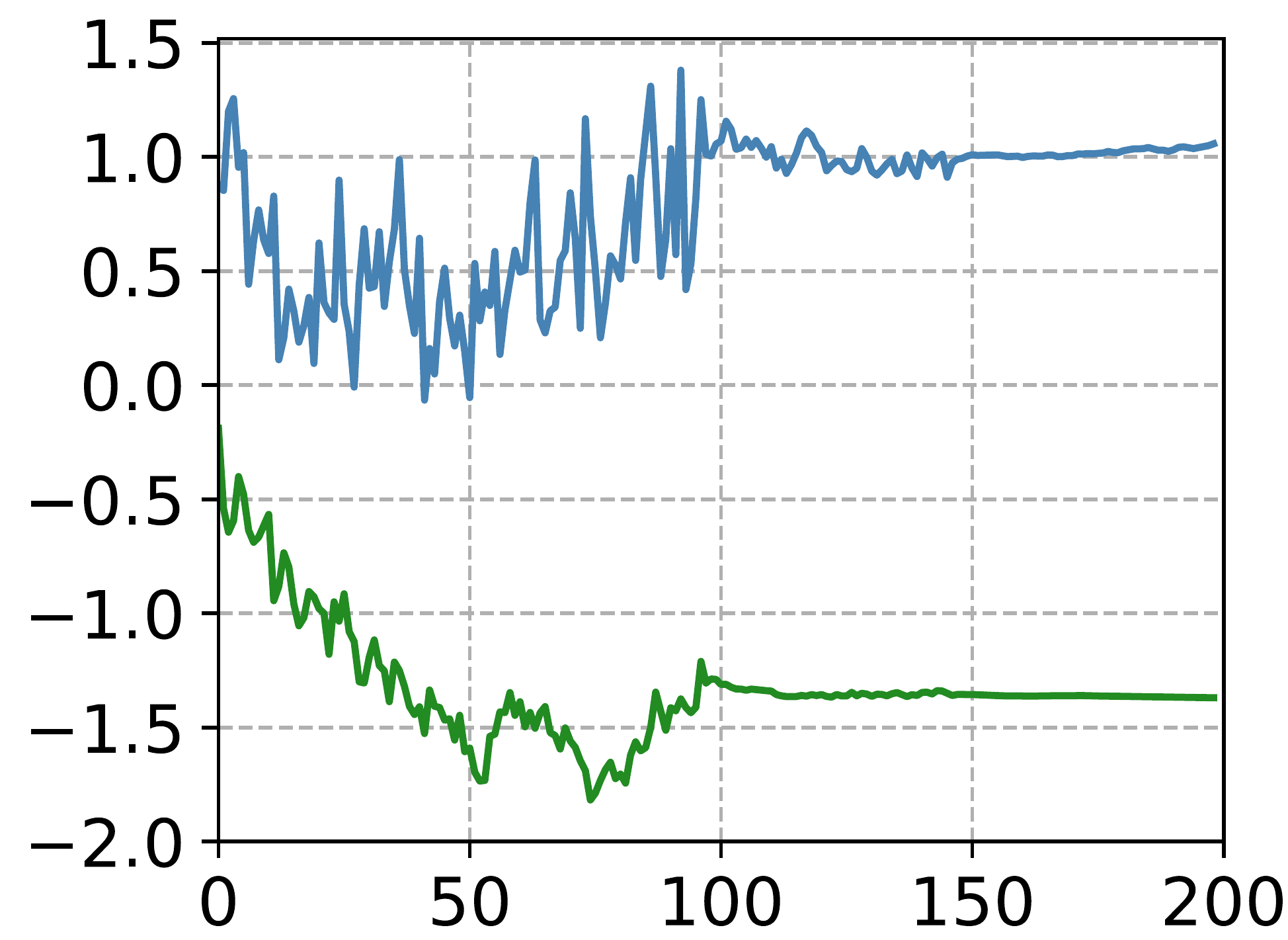}}
			\caption{Magnitude of weight update, scale factor $\gamma$ and bias term $\beta$ are having larger fluctuation as compared to passport network from scratch.}
		\end{subfigure}
		\begin{subfigure}{\textwidth}
			\centering
			{\includegraphics[keepaspectratio=true, scale = 0.25]{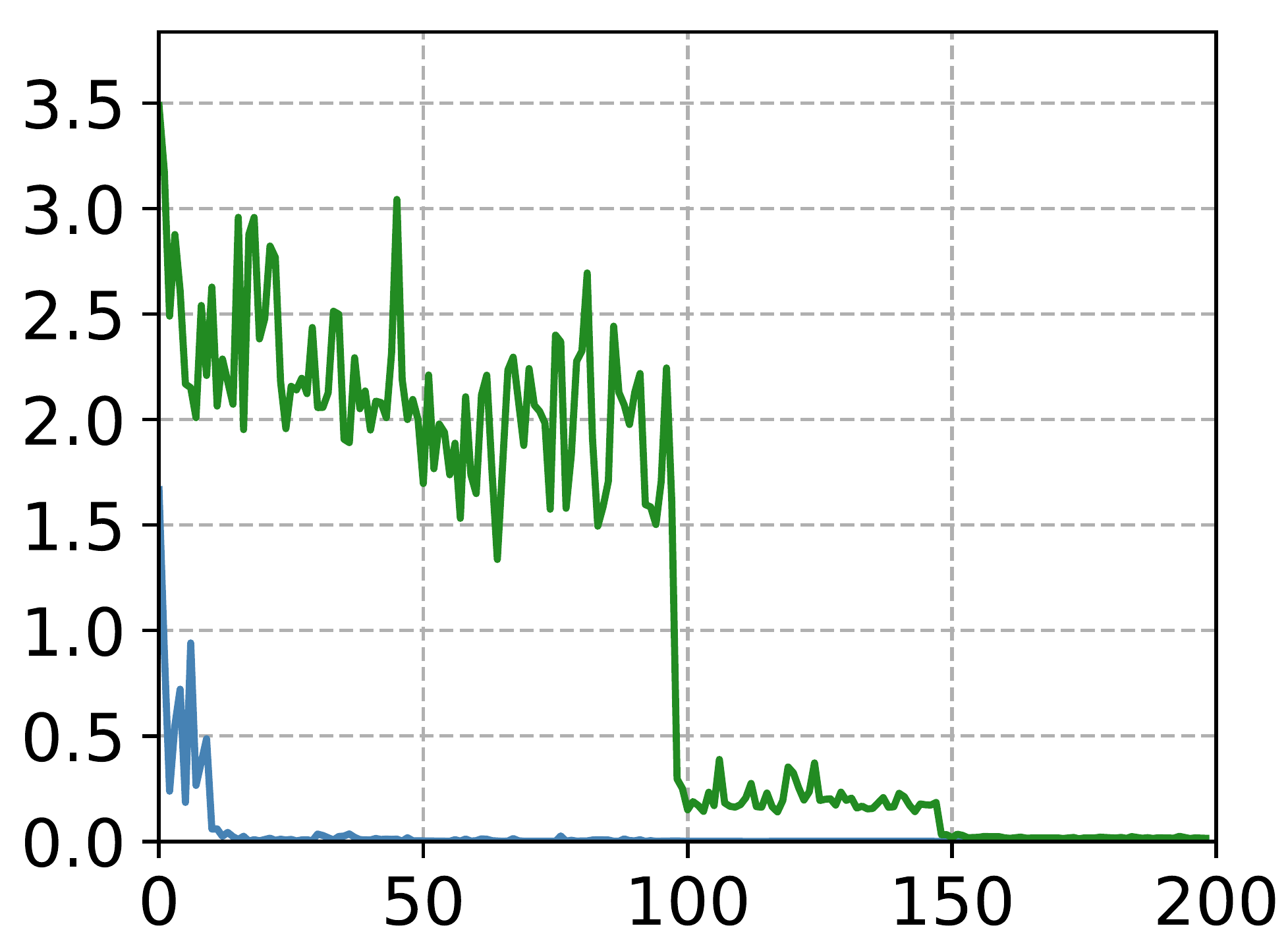}}
			{\includegraphics[keepaspectratio=true, scale = 0.25]{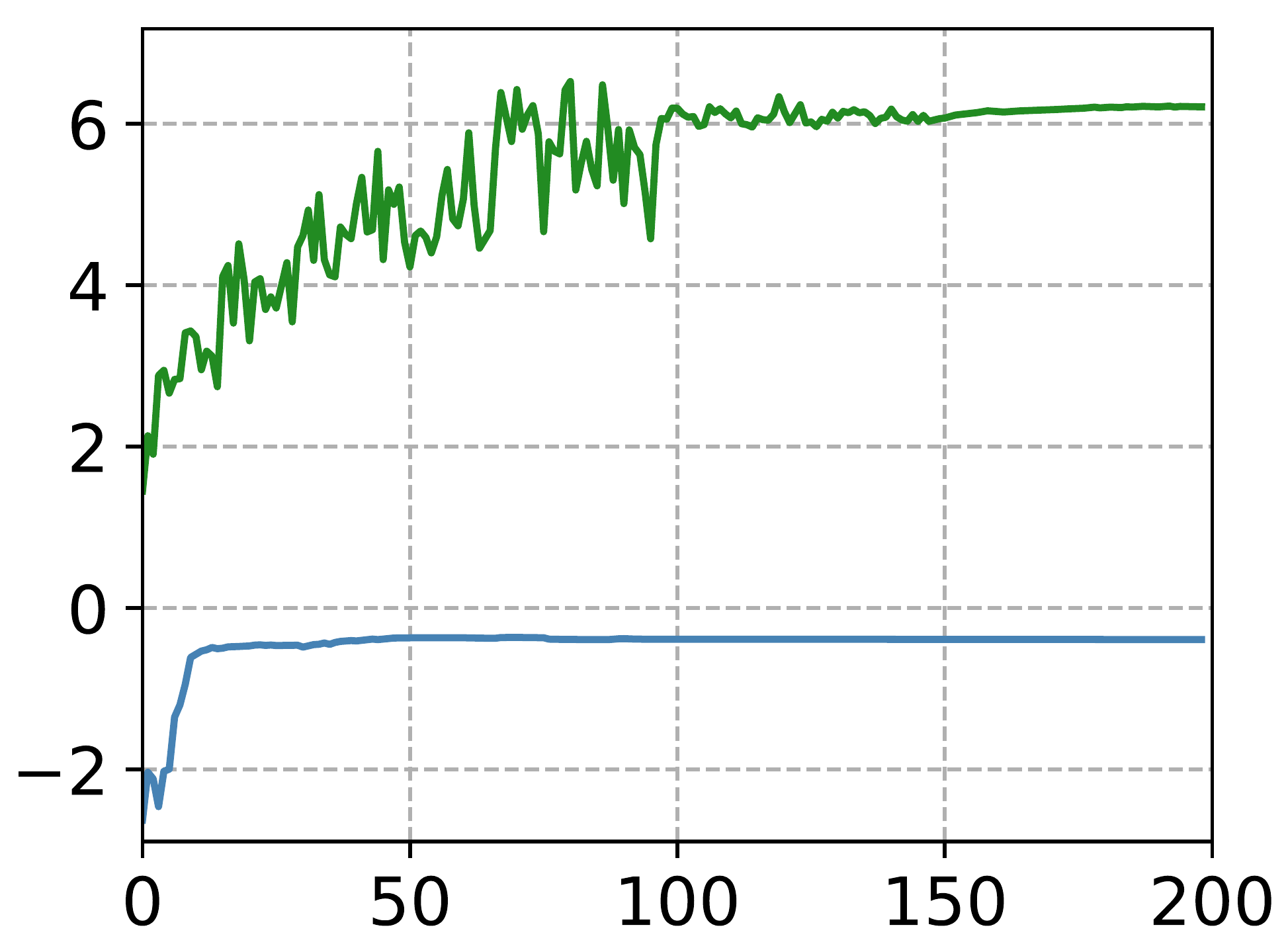}}
			{\includegraphics[keepaspectratio=true, scale = 0.25]{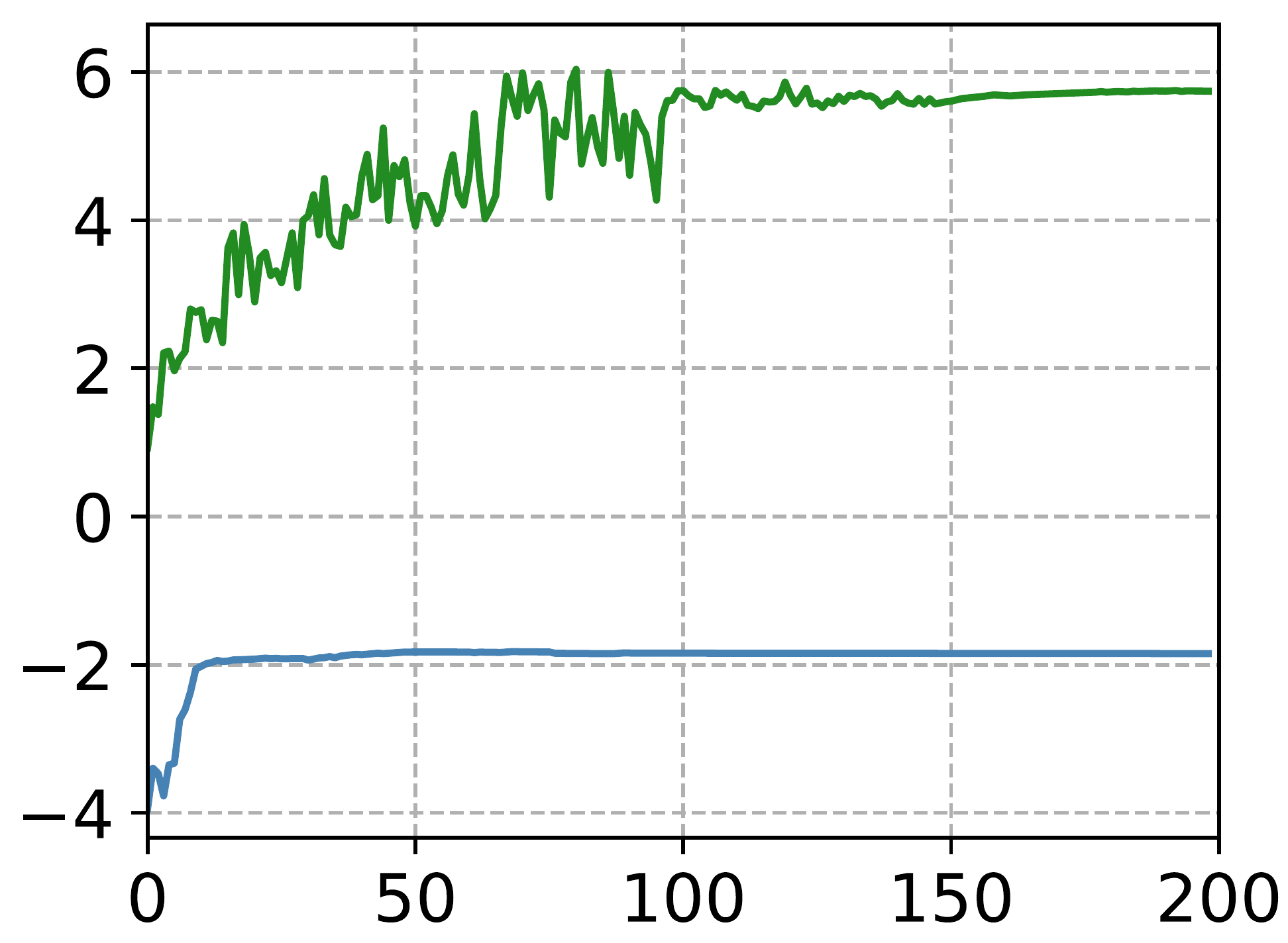}}
			\caption{Pretrained passport network \textbf{(blue)} stops update after few epochs for unknown reason}
		\end{subfigure}
		\begin{subfigure}{\textwidth}
			\centering
			{\includegraphics[keepaspectratio=true, scale = 0.25]{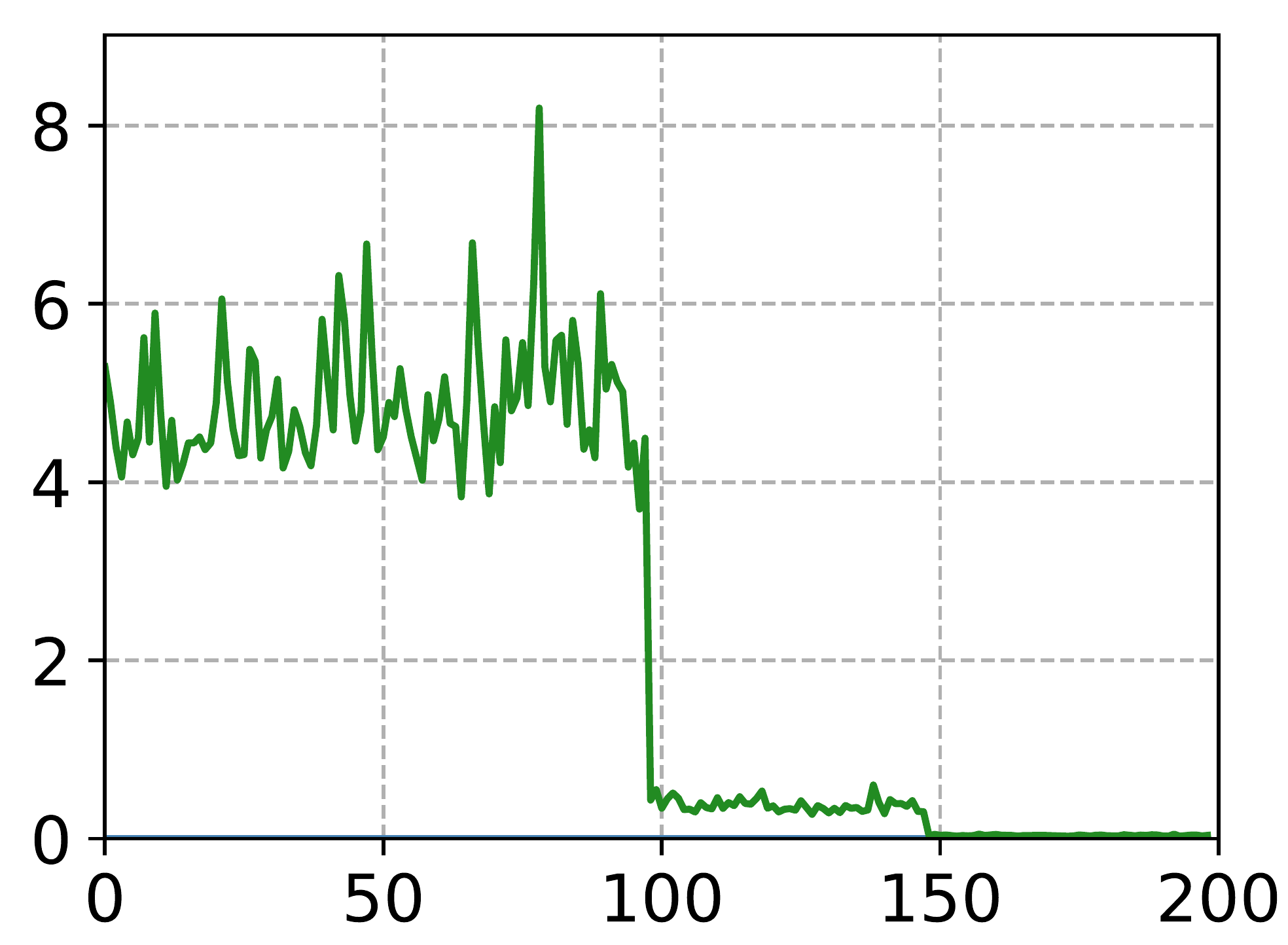}}
			{\includegraphics[keepaspectratio=true, scale = 0.25]{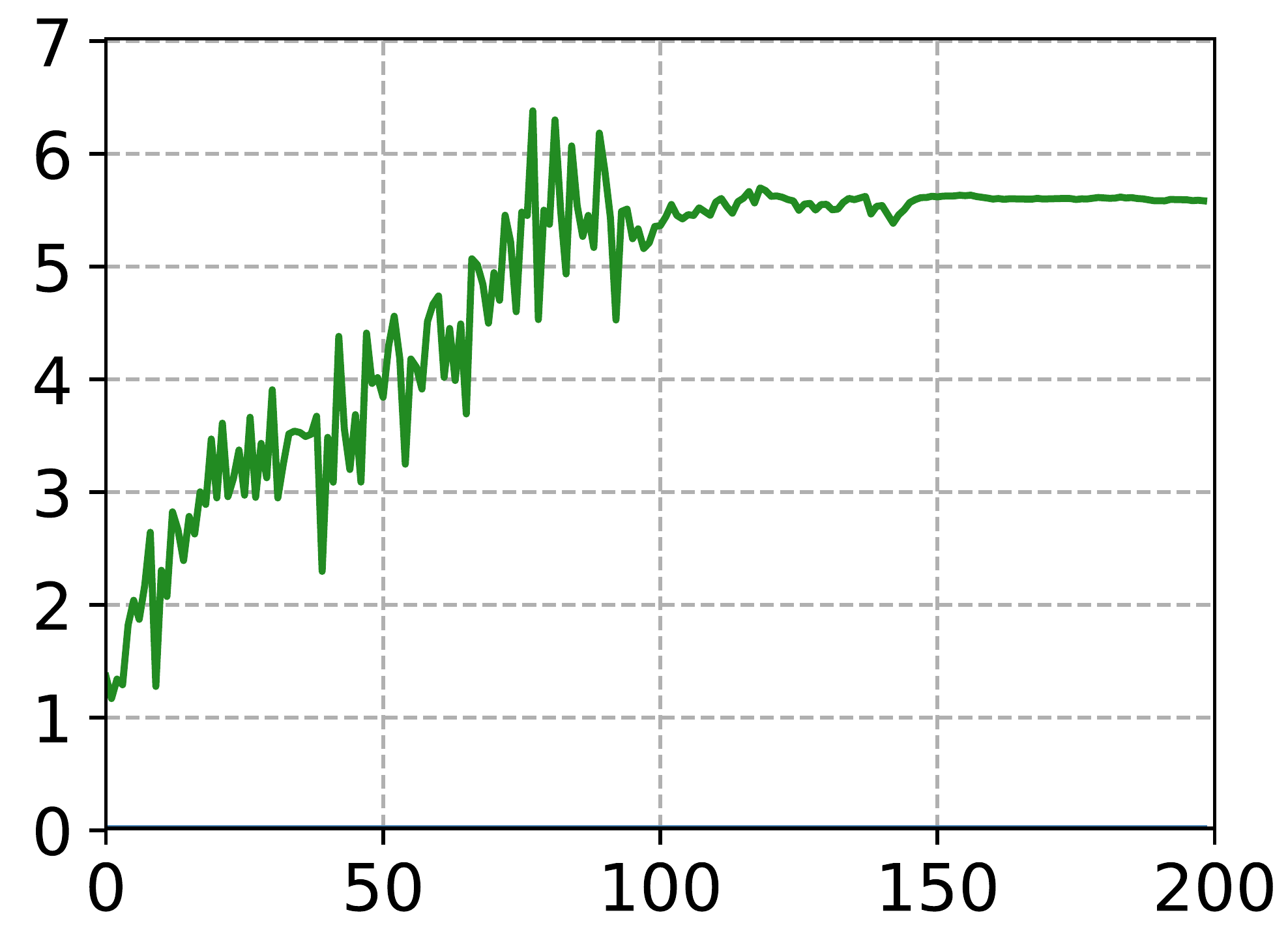}}
			{\includegraphics[keepaspectratio=true, scale = 0.25]{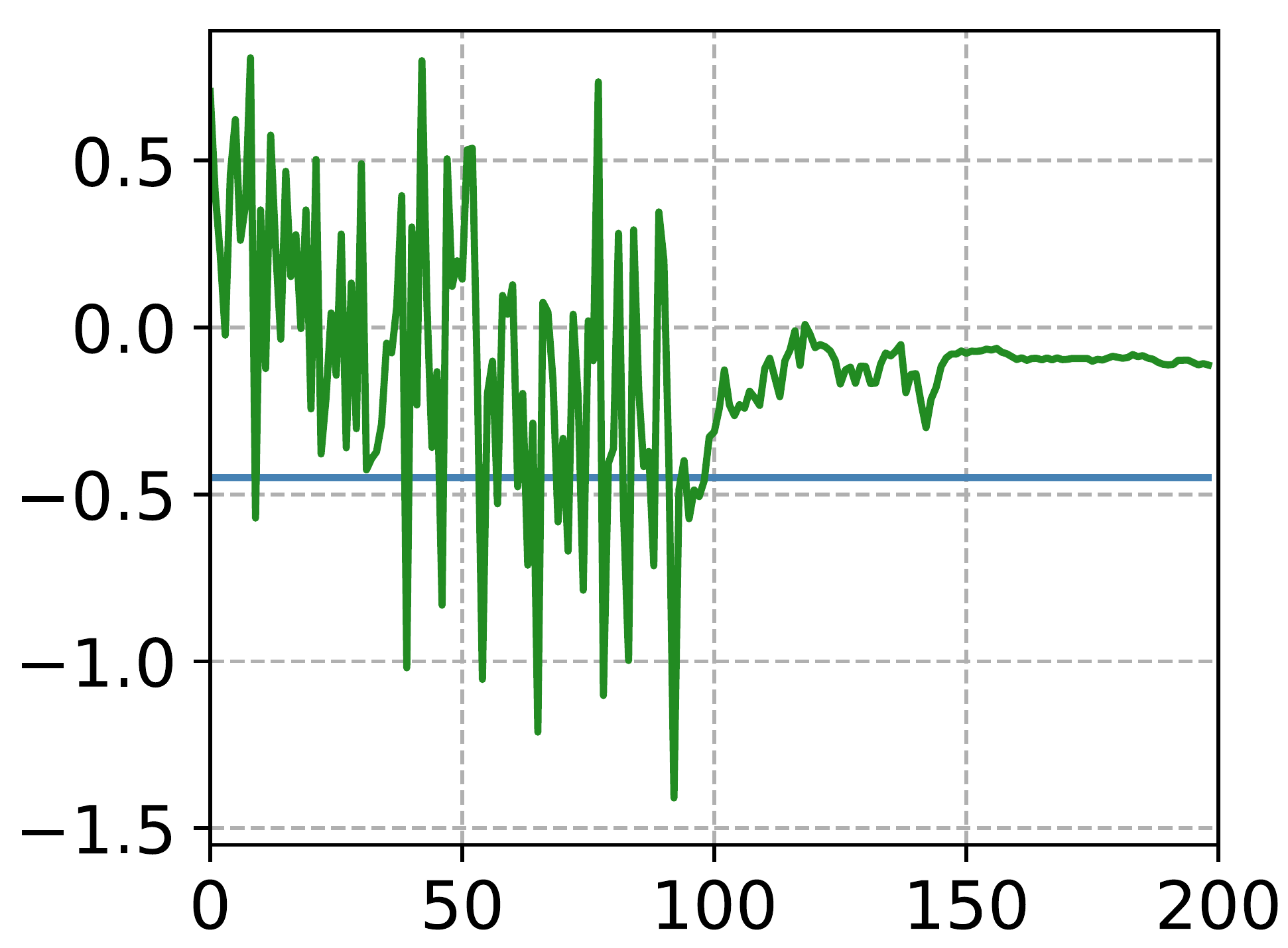}}
			\caption{Pretrained passport network \textbf{(blue)} does not update at all. Scale factor $\gamma$ stays at 0 and bias terms $\beta$ stays near -0.5}
		\end{subfigure}
		\caption{\textbf{Left to right}: Magnitudes of weight update, scale factor $\gamma$ and bias terms $\beta$ of one passporting layer over 200 epochs for the training of CIFAR10 classification. The networks are either initialized with the pre-trained model (blue) or from the scratch (green).} 
		\label{fg:oscillation}
	\end{figure*}


It turns out that the introduction of digital passporting layers does not affect the convergence of parameter tunings,
as shown in Figure \ref{fg:oscillation}, we observe that 
both the test accuracy and the computed linear transformation parameters $\gamma$ and $\beta$ stagnate in the later learning phase. 
More specifically, as demonstrated by experimental results in Section  \ref{sect:exper}, the performance discrepancies between the passport-protected networks and the original counterparts are no more than, respectively, 1\% and 5\% for cifar10 and cifar100 classification experiments investigated in this papers.
This superior \textit{functionality-preserving} capability is ascribed to the fact that the original objective functions remain unaltered during the learning stage\footnote{In contrast, the objective functions are  inevitably changed with the addition of certain regularization terms to embed watermarks \cite{EmbedWMDNN_2017arXiv,AdStitch_2017arXiv,TurnWeakStrength_Adi2018arXiv}. }.





\textbf{Digital passporting by fine-tuning:} note that the proposed digital passporting method can be applied 
with public parameters $W_p$ either being initialized from a pre-trained model or being trained from scratch using the standard initialization methods e.g. He initialization method \cite{he2015delving}. Figure \ref{fg:oscillation} shows a comparison between passport network train \textbf{from-pretrained (blue)} and \textbf{from-scratch (green)}. As shown, the training initialized with pre-trained models struggles with drastic changes in hidden parameters and the final accuracy  is inferior to that of the training from scratch approach. We therefore do not conduct more experiments with fine-tuning in our work.  




\subsection{Generation and attacking of passport} \label{sect:guess-passport}

Public parameters of a passport protected DNN might be easily plagiarized, then the plagiarizer has to deceive the network with certain passports. The chance of success of such an attacking strategy depends on the odds of correctly guessing the secret passports. 
Figure \ref{fg:samplepassport} illustrates three different types of passports which have been investigated in our work: 
\begin{itemize}
	\item[a)]  \textit{random  patterns},  whose elements are independently randomly generated according to the uniform distribution between [-1, 1]. 
	Correspondingly, the attack for this type of passports is also generated in the same vein. 
	We refer to this combination of passport-attack  as T1 type (see Table \ref{table_protection_strength_and_inconsistency}). 
	
	The chance for a random attack to coincide with the random pattern passport is extremely low, thus, strong protection against attacks are guaranteed. Yet the downside is that it is hard to associate these random patterns with person or corporate identity, which are often needed to prove the network ownership. Also, the average values of random patterns might concur with each, due to the uniform distribution of each element, thus jeopardizing the protection strengths. 
	
	\item [b)] one\footnote{Two images are needed for the V3 passport functions defined in Table \ref{tab_pass_func}.} selected image is fed through a trained network with the same architecture, and the corresponding feature maps are collected. Then the selected \textit{image} is used at the input layer and the \textit{corresponding} \textit{feature} maps are used at other layers as passports. 
	We refer to passports generated as such the \textit{fixed image} passport. The corresponding passport-attack combination is denoted as T2 in Table \ref{table_protection_strength_and_inconsistency}.
	
	The image passport is advantageous since it is straightforward to associate them with person or corporate identity. However, the protection strength provided by a single image passport is limited as plagiarizers might initiate attacks with image with similar or near duplicate contents. 
	

	\item  [c)] a set of $N$ selected \textit{images} are fed through a trained network with the same architecture, and $N$ corresponding \textit{feature maps} are collected at each layer.  Among the $N$ options only one is randomly selected as the passport at each layer. 
	Specifically, for a set of $N$ images being applied to a network with $L(<K)$ layers, there are all together $N^L$ possible combinations of passports that can be generated. 	We refer to passports generated  as such the \textit{random image} passports, which feature both strong protection strengths and easy association with person or corporate identity.

	Attackers for this type of passports have to pick up one passport at each layer, even if they have knowledge about the set of $N$ images, the chances of guessing the correct passport is merely $\frac{1}{N^L}$. 	This type of passport attacking is denoted as T3. 	
\end{itemize}
Respective performances of the above passports and attacks 
are demonstrated in Section \ref{sect:exper}.  



\section{Digital Passports as Watermarks} \label{sect:pass-as-watermark}

In case that the original training datasets are somehow made available to plagiarizers, they may opt for reverse-engineering the hidden parameters directly. The functionality of the protected networks might be retained, to various extents, depending on how successful the hidden parameters can be recovered. As shown in following subsection, the chance of success actually depends on which passport functions (in Table \ref{tab_pass_func}) are adopted.

\subsection{A reverse-engineering attack of hidden parameters} \label{subsect:reverse}

Given the original training datasets, in principle, the hidden parameters $W_h$ might be reverse-engineered by setting them as trainable variables  while holding the cloned public parameters $W_p$ as constants. Then the optimization algorithm shall adjust the hidden parameters to minimize the training error e.g. the cross-entropy loss.  

On the one hand, the reverse-engineering attack is able to successfully recover the hidden bias terms for the vulnerable choice (V1) of passport functions (with the reverse-engineered accuracies reaching almost 92\% for CIFAR10 classification). 
On the other hand, the more resilient passport functions (i.e. V2 and V3) demonstrate better \textit{protection strengths} against the reverse engineering attack --- the best accuracy the attacker can achieve is no more than 70\% (please consult more results in Section \ref{sect:expt-water} and supplementary material). Taking into account the crippled network performance, the exceedingly high computational cost as well as the high-priced skills required for setting up the reverse-engineering attack,  we regard this attacking approach unprofitable and demotivating for plagiarizers who intend to seek commercial benefits. 

\subsection{Verification of suspect network ownership}

We shall show below, even if the hidden parameters are reverse-engineered or illegally disclosed, the ownership of the protected networks can still be verified by the designed network behavior which is highly dependent on  the designated passports as formulated in (\ref{eq:dpp-formula}). Under this circumstance, digital passports play the role of digital watermarks in the white box setting \cite{EmbedWMDNN_2017arXiv}. 

As shown in Figure \ref{fg:oscillation},  the hidden parameters converge to specific \textit{constant} values $c^l_{\gamma}, c^l_{\beta}$ that lead to the desired performance i.e. $\mathcal{M}_p$. Therefore, the public parameter and the secret passport are actually constrained by the passport functions (\ref{eq:pass-gamma}) and (\ref{eq:pass-beta}):  
\begin{align}
\mathcal{F}_{\gamma}(W^l_p, p^l_{\gamma}) = c^l_{\gamma}, 
\quad \mathcal{F}_{\beta}(W^l_p, p^l_{\beta})=c^l_{\beta}. \label{eq:enforce-gamma-beta}
\end{align}

%
%

\begin{figure}[t]
	\begin{subfigure}{.3\linewidth}
		\centering
		\includegraphics[keepaspectratio=true, scale = 1]{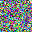}
		\caption{}
	\end{subfigure}
	\begin{subfigure}{.3\linewidth}
		\centering
		\includegraphics[keepaspectratio=true, scale = 1]{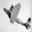}
		\caption{}
	\end{subfigure}
	\begin{subfigure}{.3\linewidth}
		\centering
		\includegraphics[keepaspectratio=true, scale = 0.14]{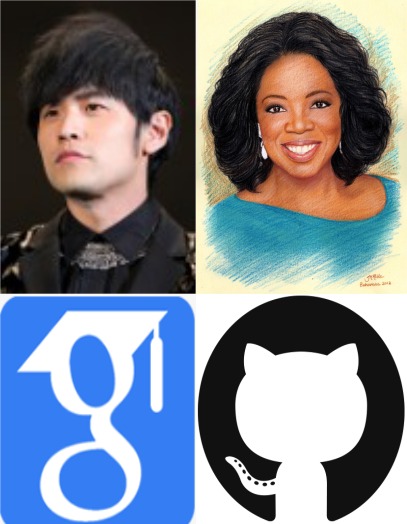}
		\caption{}
		\label{fg:logo}
	\end{subfigure}
	\caption{Example of different types of passports: (a) random patterns, (b) fixed image and (c) random image}
	\label{fg:samplepassport}
\end{figure}

\begin{figure}[t]
	\centering
		{\includegraphics[keepaspectratio=true, scale = 1]{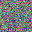}}
		{\includegraphics[keepaspectratio=true, scale = 1]{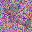}}
		{\includegraphics[keepaspectratio=true, scale = 1]{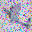}}
		{\includegraphics[keepaspectratio=true, scale = 1]{original}}
		{\includegraphics[keepaspectratio=true, scale = 0.8]{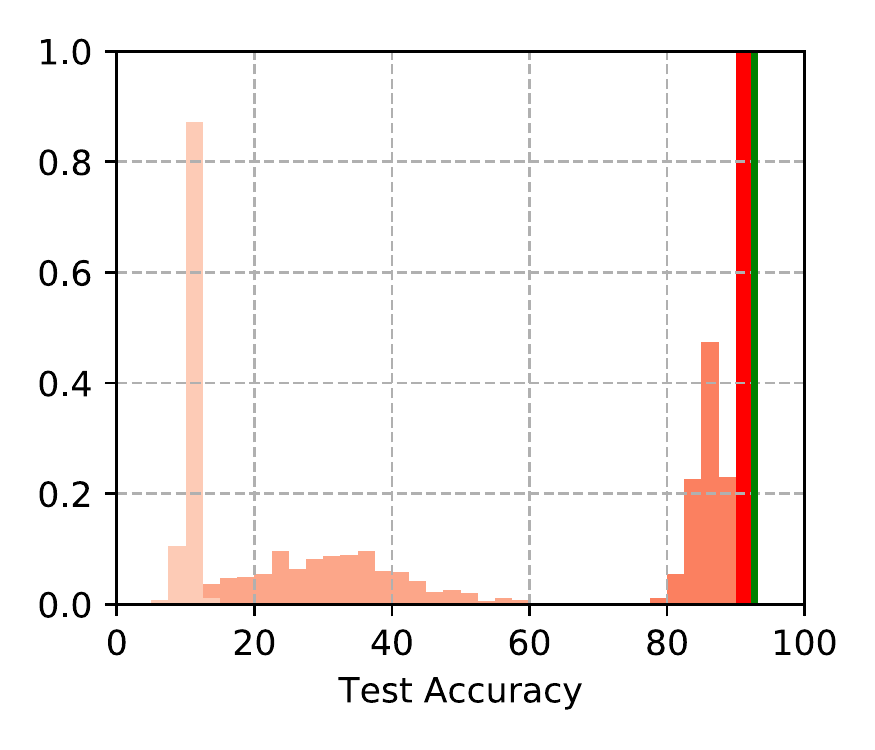}}
	\caption{Top row (right to left): the original passport image (plane) and 3 example images added with random noise to 1000, 2000, 3000 elements respectively. Bottom row: 4 red histograms (right to left) are distributions of test accuracies for CIFAR10 classification, measured with the original passport and those added with random noise as in the top row. 
}
\label{fig:noise}
\end{figure}

Bearing this constraint in mind, we propose to verify the ownership of a suspect plagiarized network by the following steps: 
\begin{enumerate}
	\item Remove the hidden variables from the network, and replace them with the passport functions  (\ref{eq:pass-gamma}) and (\ref{eq:pass-beta}) as originally designed; 
	
	\item Feed the network with the secret passports $p$ and check whether the test accuracy of a pre-determined set of test samples is the same as the expected performance $\mathcal{M}_p$.  
	Since the chance of enabling the network with a random guess is extremely low (e.g. $\frac{1}{N^L}$ for those protected by random image passport, see Section \ref{sect:guess-passport}), one can confidently claim the ownership if the verification outcome is positive;

	\item Moreover, add random noise to a varying percentage of the secret passport elements i.e. $p^e(c) = p+ e(c), c\in \{0\%, 100\%\}$, and check whether the test accuracy using passports $p^e(c)$ is the same as to the set of pre-recorded performances  $\mathcal{M}_{p^e(c)} $ (see Figure \ref{fig:noise} for an example). One can claim the ownership if the verification outcome is positive. 
	
\end{enumerate}
In order to enhance the justification of ownership, one can furthermore select either personal identification pictures or corporate logos (Figure \ref{fg:logo}) during the designing of the \textit{fixed} or \textit{random} image passports (see definitions in Section \ref{sect:guess-passport}). 

It must be noted that, using passports as proofs of ownership to stop infringements is the last resort, only if the hidden parameters are illegally disclosed or (partially) recovered. We believe this juridical protection is often not necessary since the proposed technological solution actually provides proactive, rather than reactive, IP protection of deep neural networks.




\begin{table*}[t]
	\centering \small
	\begin{tabular}{cc|c|c|c||c|c|c|}
		\cline{3-8}
		&  & \multicolumn{3}{c||}{Protection Strength ($S$ in \%)} & \multicolumn{3}{c|}{Performance Inconsistency ($I$ in \%)} \\ \hline
		\multicolumn{2}{|c|}{}   & T1 & T2 & T3 & T1 & T2 & T3 \\ \hline \hline
		\multicolumn{1}{|c|}{V1} & ResNet (92.72) & 48.39 (4.31) & 33.48 (24.21) & 76.04 (6.42) & +0.19 (0.17) & +0.21 (0.17) & +0.37 (0.18) \\ \hline \hline
		\multicolumn{1}{|c|}{V2} & ResNet (92.72) & 82.50 (0.24) & 59.54 (23.42) & 82.51 (1.05) & -0.21 (0.22) & +0.07 (0.18) & +0.10 (0.10) \\ \hline \hline
		\multicolumn{1}{|c|}{\multirow{3}{*}{V3}} & ResNet (92.72) & 82.57 (0.27) & 78.98 (8.95) & 81.86 (0.93) & -0.15 (0.20) & -0.81 (0.32) & -0.73 (0.24) \\ \cline{2-8} 
		\multicolumn{1}{|c|}{} & VGGNet (92.24) & - & - & 82.26 (0.35) & - & - & +0.02 (0.26) \\ \cline{2-8} 
		\multicolumn{1}{|c|}{} & AlexNet (86.41) & - & - & 76.83 (1.59) & - & - & +0.82 (0.23) \\ \hline
	\end{tabular}
	\caption{ A comparison of CIFAR10 classification  performances, in terms of \textit{protection strength} and \textit{performance inconsistency} of protected networks against various attacks. 
	The scores next to each network are test accuracies  of the original networks ($A_o$). Other scores are average $S$ or $I$, and in bracket() are standard deviations.}
	\label{table_protection_strength_and_inconsistency}
\end{table*}

\section{Experiments} \label{sect:exper}

\subsection{Experiment setup}

In our experiments, we investigated two image classification tasks i.e. CIFAR10 and CIFAR100, with three popular deep learning architectures i.e. ResNet \cite{he2016deep}, VGGNet \cite{simonyan2014very} and the seminal Alexnet \cite{krizhevsky2012imagenet}. Detailed descriptions of the datasets, network architectures as well as the hyperparameters of the training algorithm are elaborated in the supplementary material. 

\begin{figure*}[t]
	\centering
	 \begin{subfigure}{\textwidth}
	 \centering
	{\includegraphics[keepaspectratio=true, scale = 0.6]{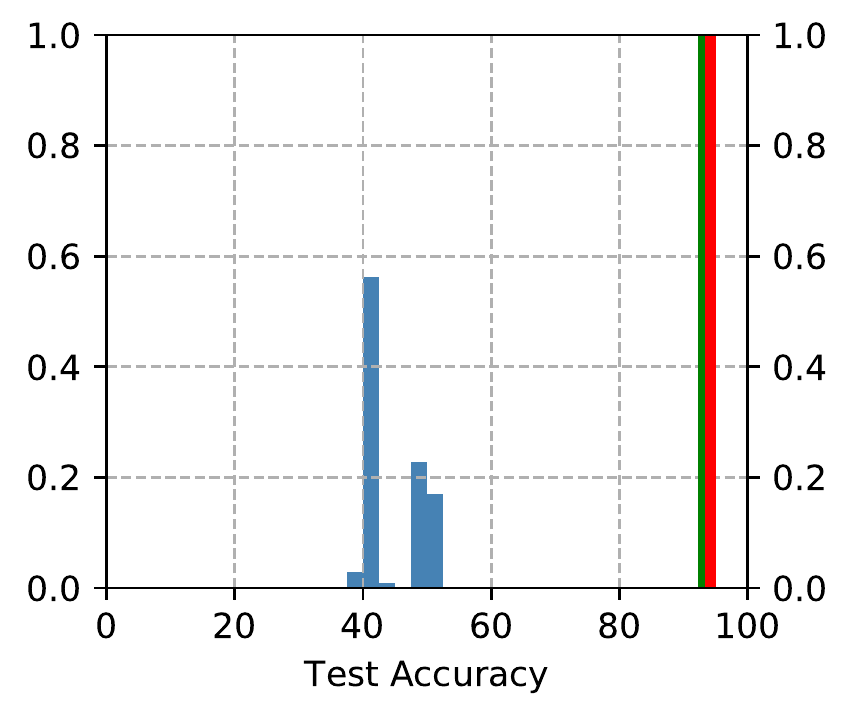}}
	{\includegraphics[keepaspectratio=true, scale = 0.6]{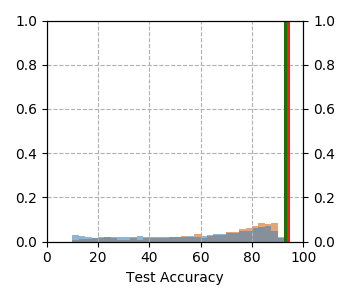}}
	{\includegraphics[keepaspectratio=true, scale = 0.6]{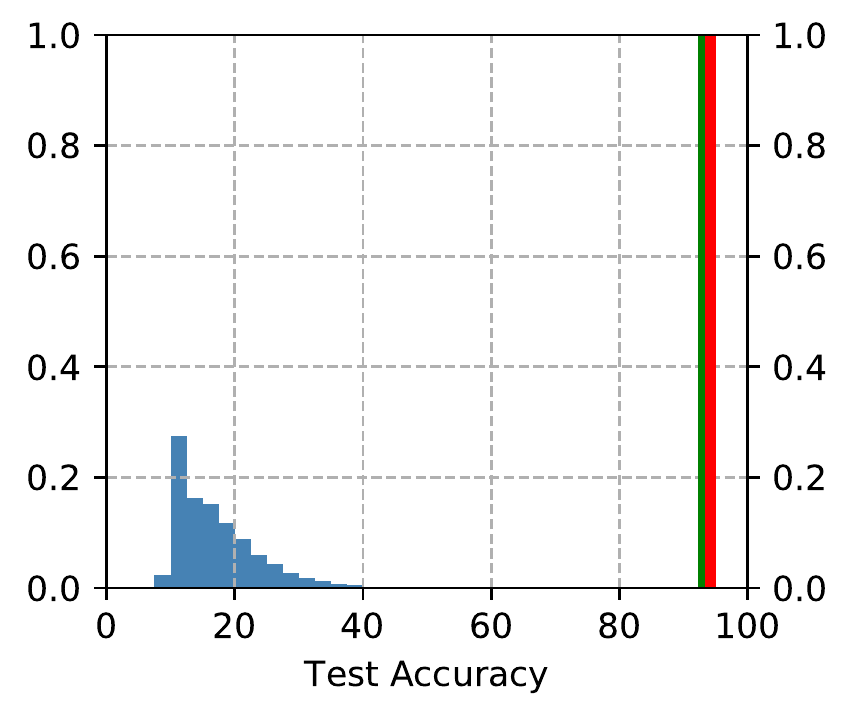}}
	\caption{Passport function V1}
	\end{subfigure}

\begin{subfigure}{\textwidth}
	 \centering
	{\includegraphics[keepaspectratio=true, scale = 0.6]{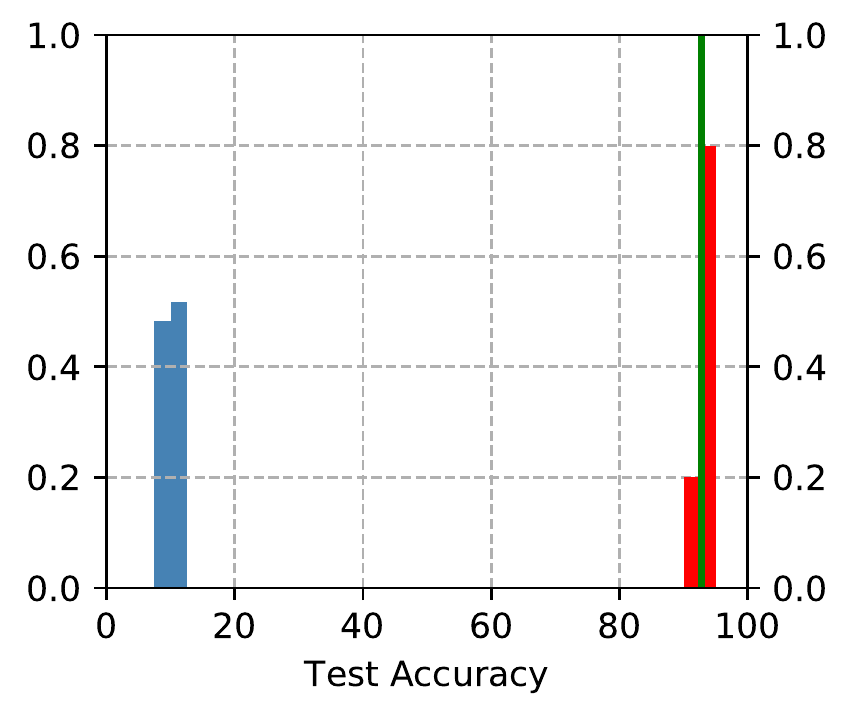}}
	{\includegraphics[keepaspectratio=true, scale = 0.6]{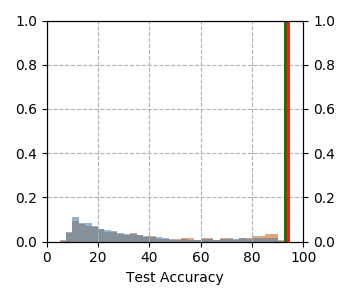}}
	{\includegraphics[keepaspectratio=true, scale = 0.6]{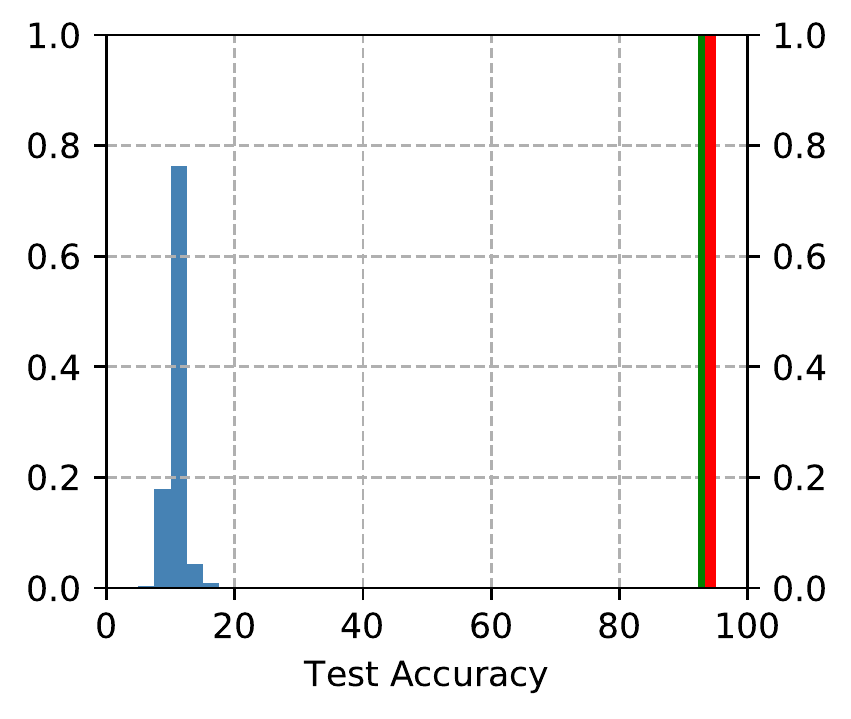}}
	\caption{Passport function V2}
	\end{subfigure}

\begin{subfigure}{\textwidth}
	 \centering
	{\includegraphics[keepaspectratio=true, scale = 0.6]{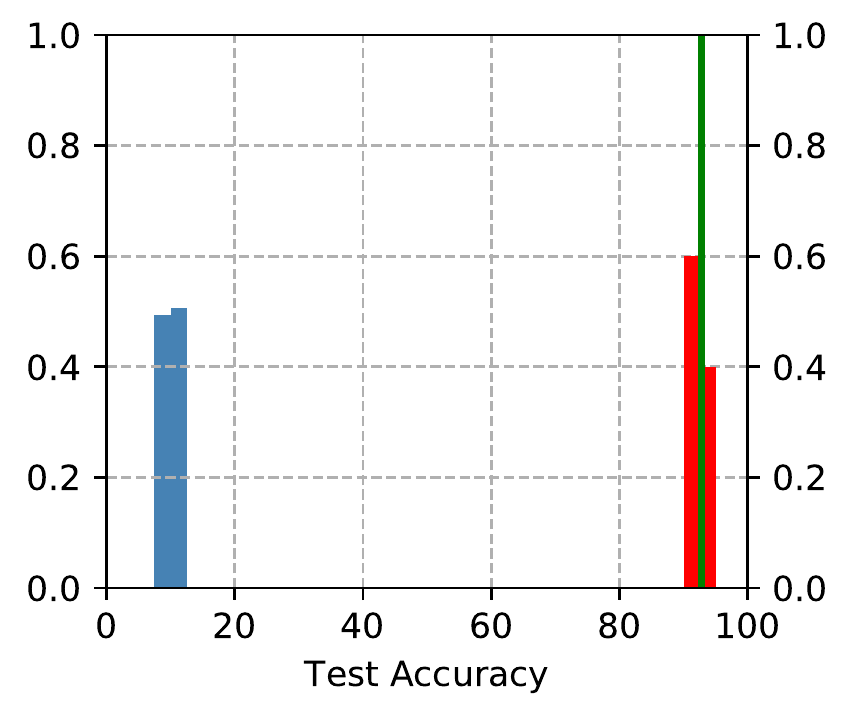}}
	{\includegraphics[keepaspectratio=true, scale = 0.6]{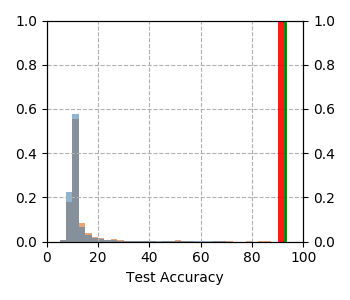}}
	{\includegraphics[keepaspectratio=true, scale = 0.6]{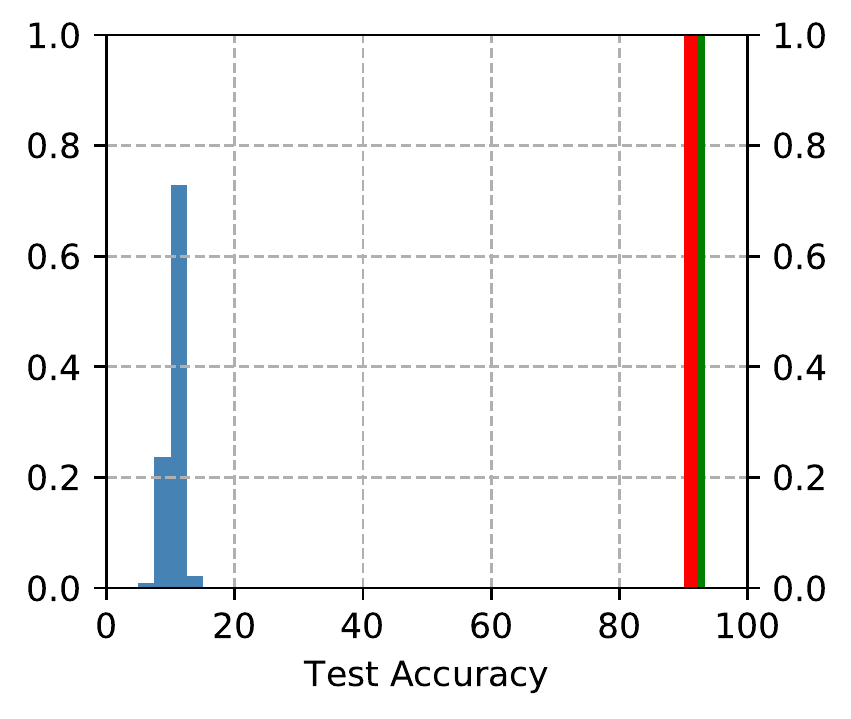}}
	\caption{Passport function V3}
	\end{subfigure}
	
	\caption{CIFAR10: Distributions of the test accuracies (\%) with \textit{ResNet}.  
		\textbf{Each histogram}: green line (test accuracy of the original network), red histogram (test accuracies with the correct passports) and blue histogram (test accuracies with fake passports).  
		\textbf{Left to right}: Passport types as described at Section \ref{sect:guess-passport} - random patterns, fixed image and random image passports, respectively. 
	}
	\label{fig:hist-resnet-cifar10}
\end{figure*}

For each task \& datasets, we trained multiple networks with different passport functions and tested them against different attacking strategies. Performances of the passport-protected network were reported using histograms of their respective accuracies (see Figure \ref{fig:hist-resnet-cifar10} for the experiment results). In terms of quantitative evaluation metrics, we adopted the \textit{performance inconsistency} and \textit{protection strength}, defined in Section \ref{sect:formulation}, which are also given as follows: 
\begin{align}
	I = A_{o} - A_{p}, \quad
	S = A_{p} - A_{t},
\end{align} in which $I$ stands for inconsistency, $S$ for protection strength, and  $A_{o}, A_{p}, A_{t}$ denote, respectively, test accuracies of the original, the protected and the attacked networks. 

\begin{figure*}[h]
		\begin{subfigure}{\textwidth}
			\centering
			{\includegraphics[keepaspectratio=true, scale = 0.6]{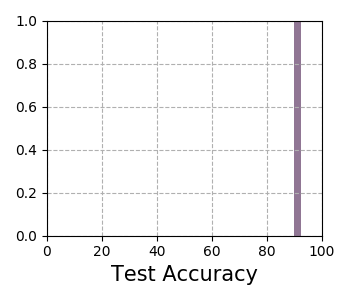}}
			{\includegraphics[keepaspectratio=true, scale = 0.6]{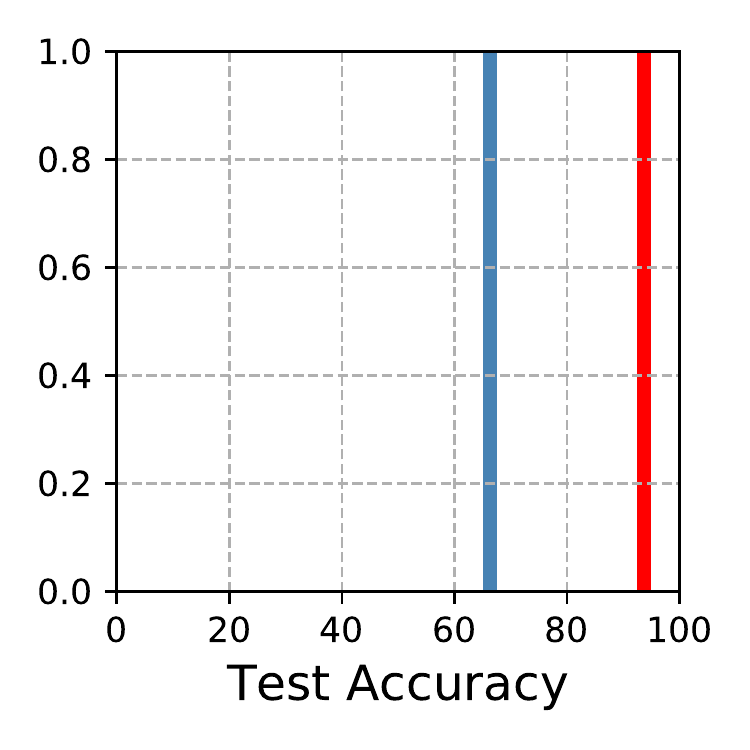}}
			{\includegraphics[keepaspectratio=true, scale = 0.6]{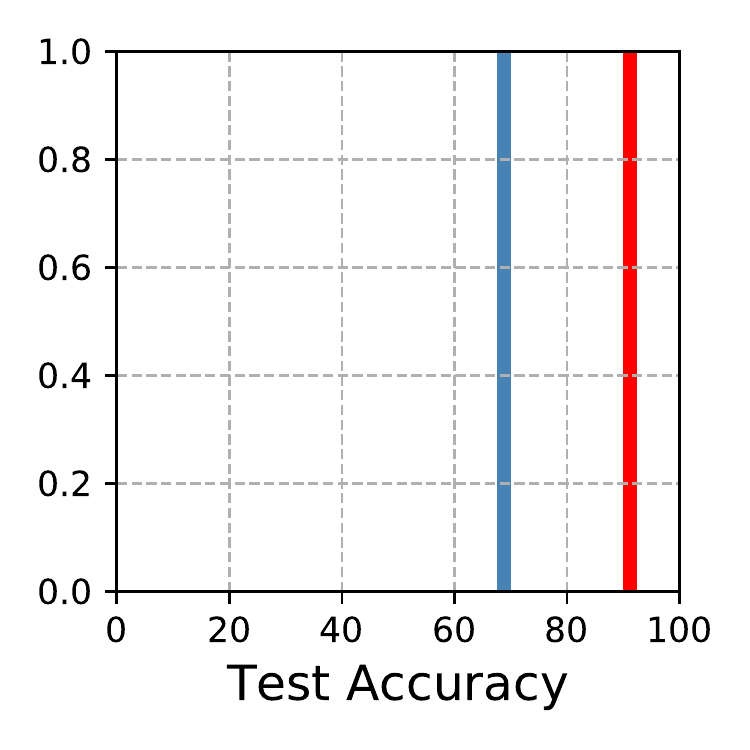}}
			\caption{ResNet18}
		\end{subfigure}
		
		\begin{subfigure}{\textwidth}
			\centering
			{\includegraphics[keepaspectratio=true, scale = 0.6]{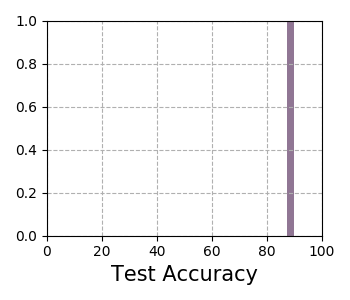}}
			{\includegraphics[keepaspectratio=true, scale = 0.6]{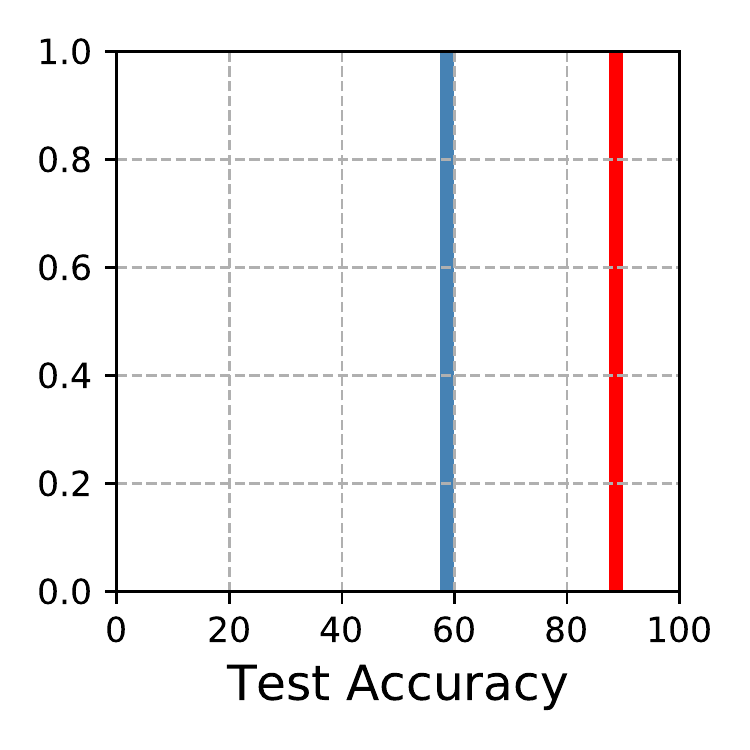}}
			{\includegraphics[keepaspectratio=true, scale = 0.6]{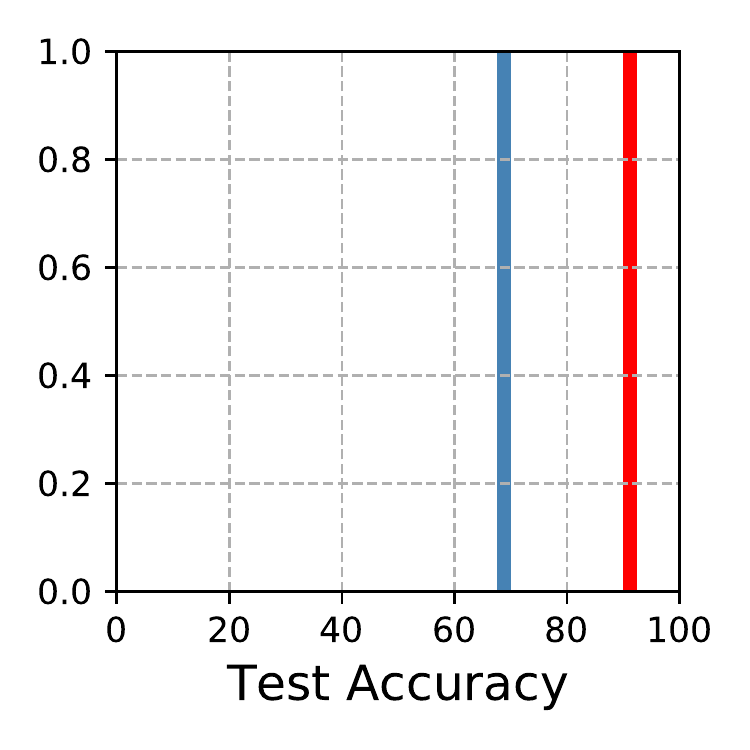}}
			\caption{VGG16}
		\end{subfigure}
		
		\begin{subfigure}{\textwidth}
			\centering
			{\includegraphics[keepaspectratio=true, scale = 0.6]{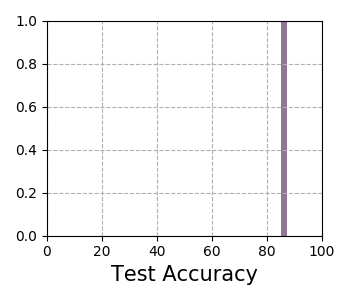}}
			{\includegraphics[keepaspectratio=true, scale = 0.6]{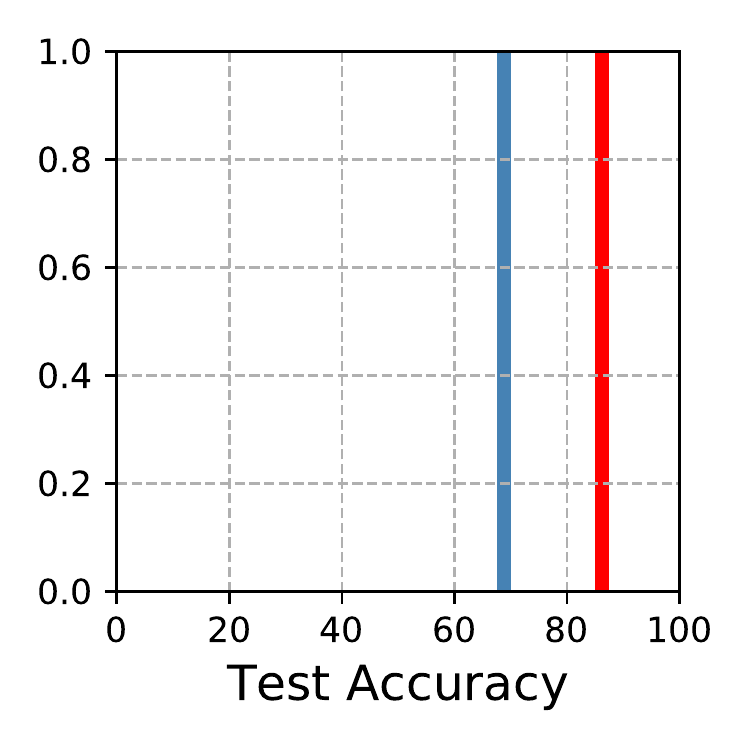}}
			{\includegraphics[keepaspectratio=true, scale = 0.6]{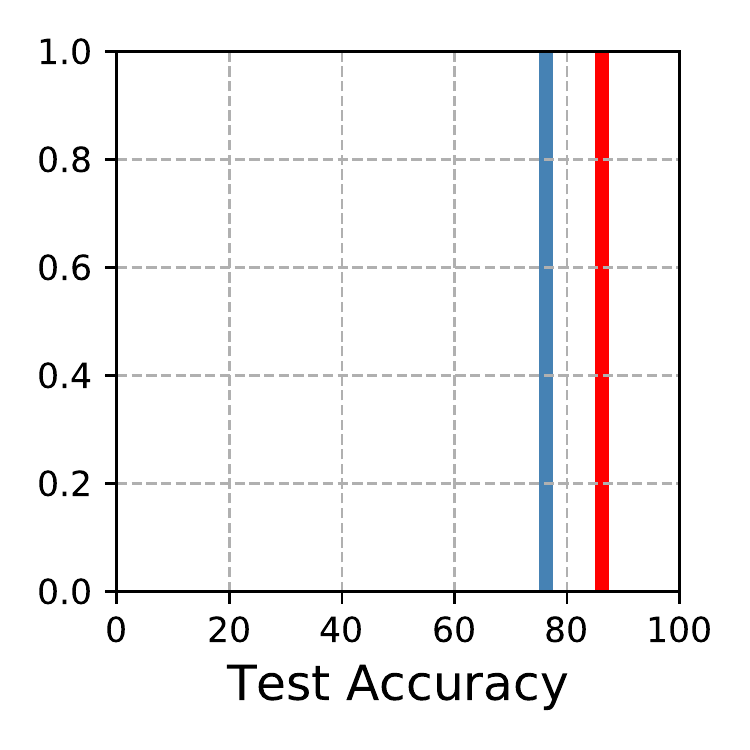}}
			\caption{AlexNet}
		\end{subfigure}
		
		\caption{Histogram of reverse engineering attack for passport type \textbf{T3}: \textbf{random images}. \textbf{Left to right}: Passport function V1 to V3. Due to histogram bar overlapping, the purple colour bar is the resultant colour of blue colour overlaps with red colour.}
\label{figurerev}
	\end{figure*}
	
		\begin{table}[t]
	\centering
	\resizebox{0.45\textwidth}{!}{ 
	\begin{tabular}{cC{3cm}|c|c||c|c|}
		\cline{3-4}
		&  & Protection  & Performance  \\
		&  & Strength  & Inconsistency  \\ \hline 
	\hline
		\multicolumn{1}{|c|}{\multirow{3}{*}{V3}} & ResNet (70.19) & 71.14 (0.19) & +1.99 (0.23) \\ \cline{2-4}
		\multicolumn{1}{|c|}{} & VGGNet (70.86) & 66.98 (0.01) & -2.88 (0.37) \\ \cline{2-4} \cline{2-4}
		\multicolumn{1}{|c|}{} & AlexNet (58.19) & 61.33 (0.33) & +4.21 (0.49) \\ \hline
	\end{tabular}}
	\caption{CIFAR100: A comparison in terms of \textit{protection strength} and \textit{performance inconsistency} of various protected networks against T3 attacks. 	The scores next to each network are test accuracies of the original networks ($A_o$). Other scores are average $S$ or $I$, and in bracket() are standard deviations.}
	\label{table_cifar100}
\end{table}

\subsection{Protection performances against passport attacks}\label{sect:expt-pass}

Following experiments are carried out to evaluate the performance of the proposed digital passport method. First, for any given network architectures, they are embedded with the three different types of passports introduced in Section \ref{sect:guess-passport} where each network repeats 5 times with different passports being embedded. Test accuracies of the 5 protected networks with corresponding valid passports are measured.  Second, for each passport embedded network, 1000 fake-passport attacks are attempted with resulting test accuracies measured.
Third, for each protected network, three different passport functions (i.e. V1,V2,V3) introduced in Table \ref{tab_pass_func} are adopted and the resulting test accuracies are measured. 

Figure \ref{fig:hist-resnet-cifar10} illustrates histograms of CIFAR10 classification accuracies measured, respectively, for the original network, the protected network with valid or fake passports, using different passport functions. 
Table \ref{table_protection_strength_and_inconsistency} summarizes the averaged performance inconsistencies and protection strengths, over all passport protected networks.  

{\bf Functionality-preserving:} First, it is observed that the performance inconsistency between the original and the protected networks is no more than 1\% for all networks. This is an important result as it shows that the original objective functions remain unaltered during the learning stage. 

{\bf Well-protected:} Second, it shows that the protection strength of the digital passport based network is ranging from 33\% up to 83\%. Among three types of passport functions, the V3 passport function provides the most resilient protection with the protection strength consistently larger than 76\%. 

As a summary, the networks embedded with passport functions V2/V3 
are consistently {\it functionality-preserving} and {\it well-protected} when $\tau_d=1\%, \tau_s=50\%$, as defined in Section \ref{sect:formulation}. In contrast, the combination of V1 passport functions and the fixed image passport provides the most vulnerable protection, with averaged protection strength merely being 33\%. Inspection of the histogram in Figure \ref{fig:hist-resnet-cifar10} shows that the most aggressive attack may achieve the test accuracy near 90\%. As a result of this vulnerability, for the experiments with other network architectures and CIFAR100, we skip the V1 and only use the V3 passport function, together with the random image passports and T3 passport attacks.


{\bf CIFAR100: } We also conducted the same experiments in this public dataset (see Table \ref{table_cifar100} for results), and the performance inconsistency is between -3 to 4\% and the network protection strength is ranging from 61\% up to 71\%. Note that for ResNet and Alexnet, the test accuracies of the protected networks are actually higher than the original. Also, the fake-passport attacking accuracies for both ResNet and VGGNet are about 1\%, virtually equivalent to random guessing of 1 out 100 classes. 

\begin{table}[t]
	\centering
	\resizebox{0.47\textwidth}{!}{ 
	\begin{tabular}{cC{3.5cm}|C{1cm}|C{1cm}||C{1cm}|C{1cm}|}
		\cline{3-6}
		&  & \multicolumn{2}{c||}{Protection} & \multicolumn{2}{c|}{Performance } \\
		&  & \multicolumn{2}{c||}{Strength} & \multicolumn{2}{c|}{Inconsistency} \\ \hline
		\multicolumn{2}{|c|}{}   & T1  & T3 & T1  & T3 \\ \hline \hline
		\multicolumn{1}{|c|}{\multirow{3}{*}{V1}} & ResNet (92.72) & 0.56 & 0.19 & -0.26 & -0.26 \\ \cline{2-6} 
		\multicolumn{1}{|c|}{} & VGGNet (92.24) & 0.13 & 0.18 & +2.7 & +2.64 \\ \cline{2-6} 
		\multicolumn{1}{|c|}{} & AlexNet (86.41) & 0.24 & 0.15 & -1.38 & -1.58\\ \hline \hline
		\multicolumn{1}{|c|}{\multirow{3}{*}{V2}} & ResNet (92.72) & 26.61 & 25.36 & +0.02 & +0.03\\ \cline{2-6} 
		\multicolumn{1}{|c|}{} & VGGNet (92.24) & 33.57 & 31.55 & +3.56 & +2.55\\ \cline{2-6} 
		\multicolumn{1}{|c|}{} & AlexNet (86.41) & 19.24 & 18.02 & -0.15 & -0.93\\ \hline \hline
		\multicolumn{1}{|c|}{\multirow{3}{*}{V3}} & ResNet (92.72) & 25.69 & 22.93 & +0.31 & +0.31 \\ \cline{2-6} 
		\multicolumn{1}{|c|}{} & VGGNet (92.24) & 33.36 & 22.21 & +2.77 & +0.36 \\ \cline{2-6} 
		\multicolumn{1}{|c|}{} & AlexNet (86.41) & 14.85 & 12.00 & -1.25 & -0.88 \\ \hline
	\end{tabular}}
	\caption{A comparison of CIFAR10 classification performances, in terms of \textit{protection strength} and \textit{performance inconsistency} of protected networks against RevAs attacks. The scores next to each network are test accuracies of the original networks ($A_o$).}\vspace{-10pt}
	\label{reve_table}
\end{table}

\subsection{Protection performance against reverse-engineering attacks} \label{sect:expt-water}

For each network model constructed for experiments in Section \ref{sect:expt-pass}, we apply the reverse-engineering attack (RevA) as illustrated in Section \ref{subsect:reverse} and measure the performance of the recovered networks.  Each type of network repeats 5 times as stated before, and Figure \ref{figurerev} illustrates the histograms of measured accuracies distributions, respectively, for the reverse-engineered networks and the protected network with valid passports. 
Corresponding protection strengths and performance inconsistency are summarized in Table \ref{reve_table}. 

It was observed that the network protected by the V1 passport function is vulnerable to RevAs, and this is particularly true for the random image passport.  On the other hand, the networks have a better protection with V2/V3 passport functions, where the protection strengths are around 24\%.  While the network functionality is not completely disabled, taking into account the high computational costs of RevAs attacks, we view the protection against RevAs is effective. 

Finally, as a last resort, the embedded passport images can be associated with person or corporate identities as depicted in Figure \ref{fg:logo}. This provides an easily verifiable approach to claim ownership of protected networks. 

\subsection{Distribution of test accuracies with different deep architecture}
	
\begin{figure*}[ht]
		\centering
		\begin{subfigure}{.45\linewidth}
			\centering
			\includegraphics[keepaspectratio=true, scale = 0.35]{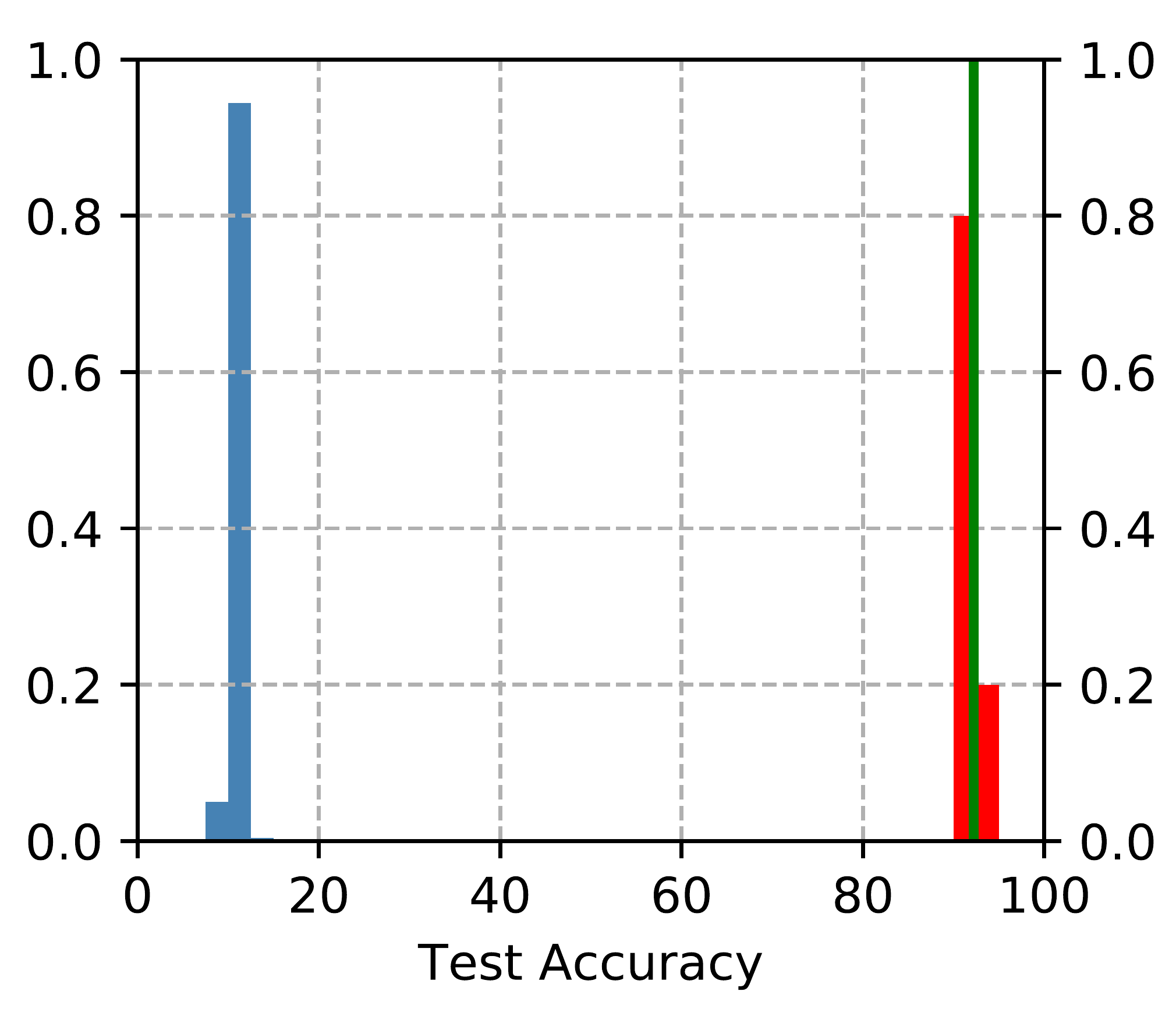}
			\caption{VGGNet}
		\end{subfigure}
		\begin{subfigure}{.45\linewidth}
			\centering
			\includegraphics[keepaspectratio=true, scale = 0.35]{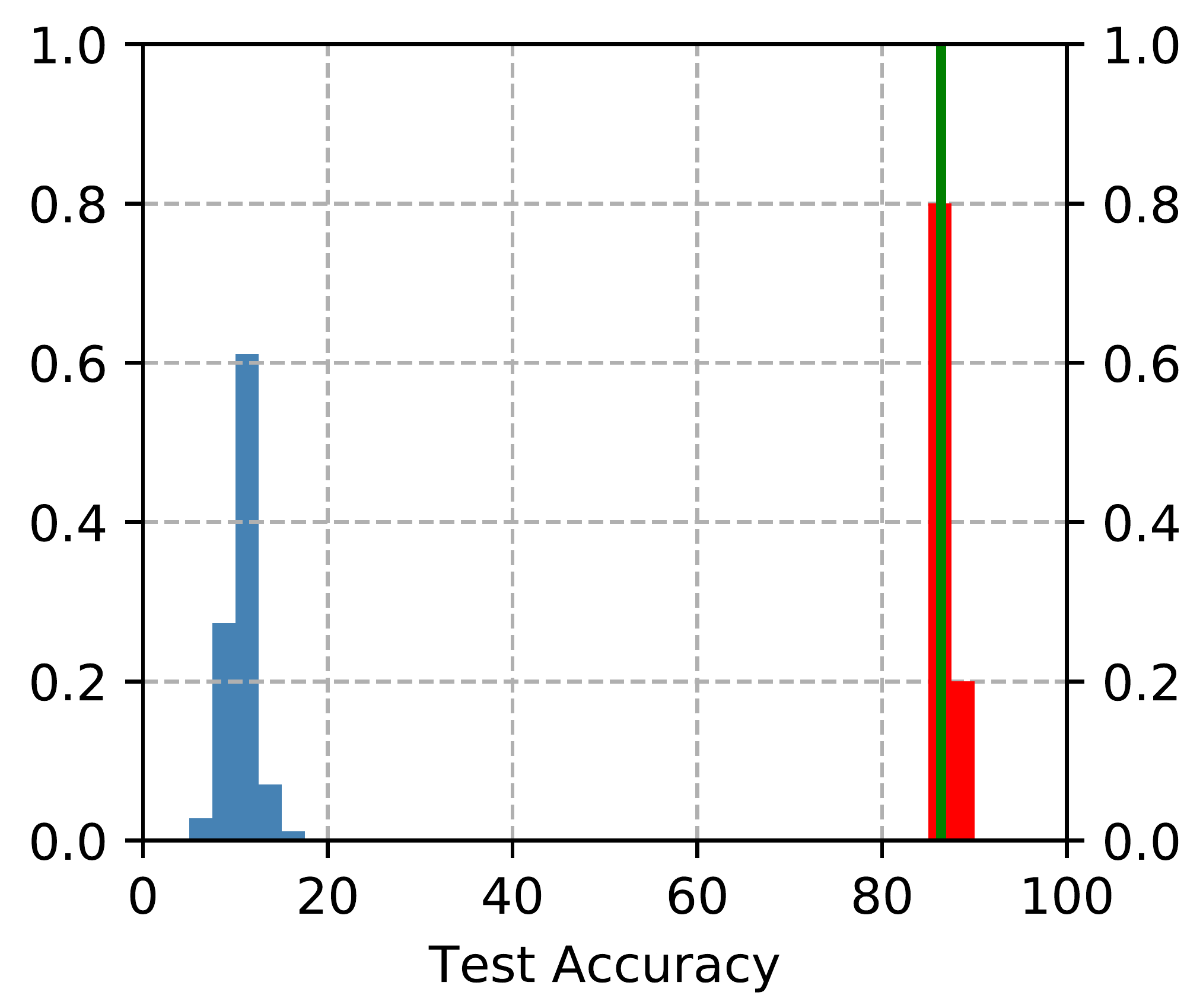}
			\caption{AlexNet}
		\end{subfigure}
		\caption{CIFAR10: Distributions of the test accuracies (\%) with \textit{VGGNet} and \textit{AlexNet} using passport function V3 with random image passports.
		} \vspace{-10pt}
		\label{fig:hist-alexnet-vgg16-cifar10} 
	\end{figure*}

\begin{figure*}[ht]
		\centering
		\begin{subfigure}{.3\linewidth}
			\centering
			\includegraphics[keepaspectratio=true, scale = 0.25]{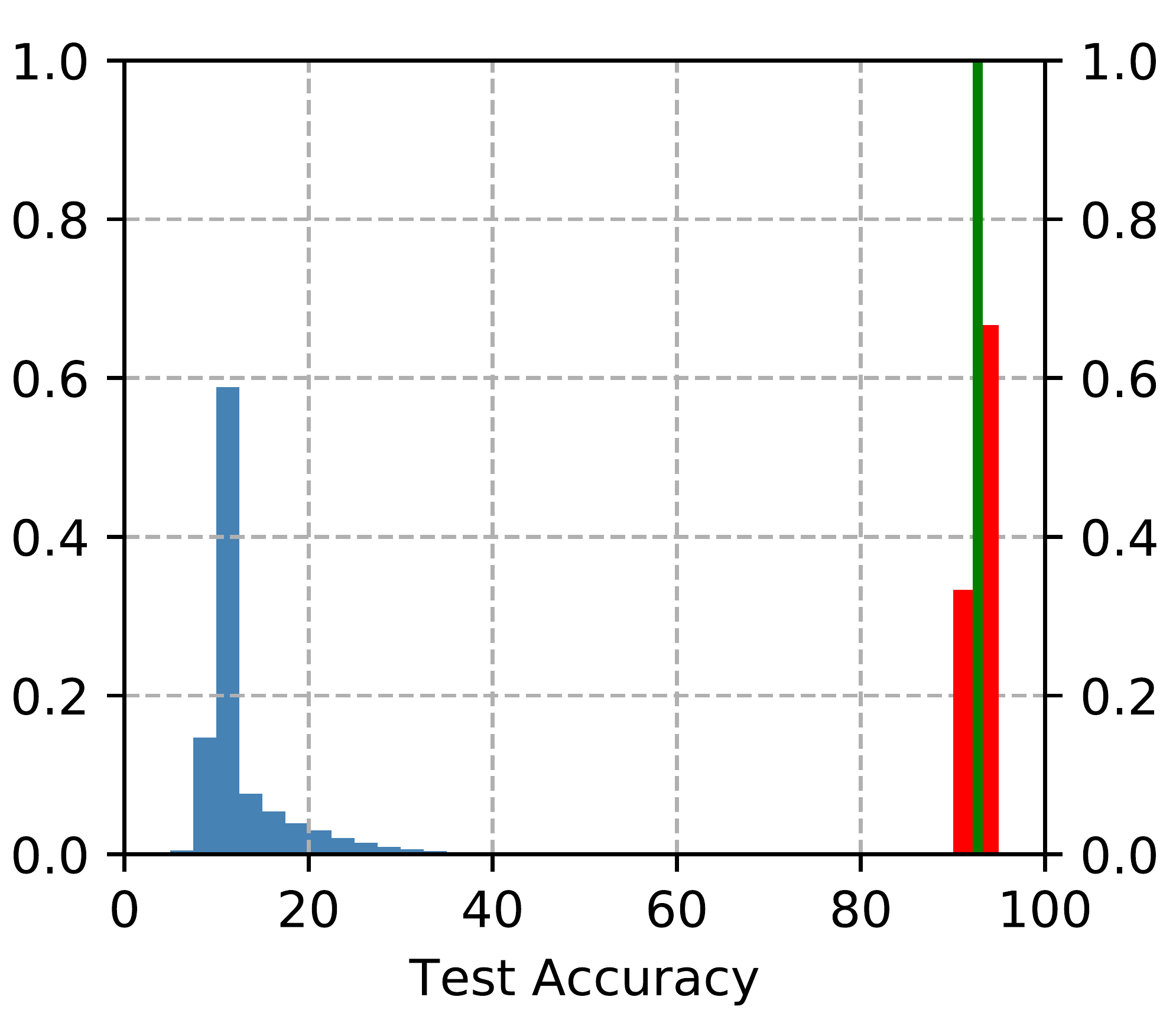}
			\caption{ResNet}
		\end{subfigure}
		\begin{subfigure}{.3\linewidth}
			\centering
			\includegraphics[keepaspectratio=true, scale = 0.25]{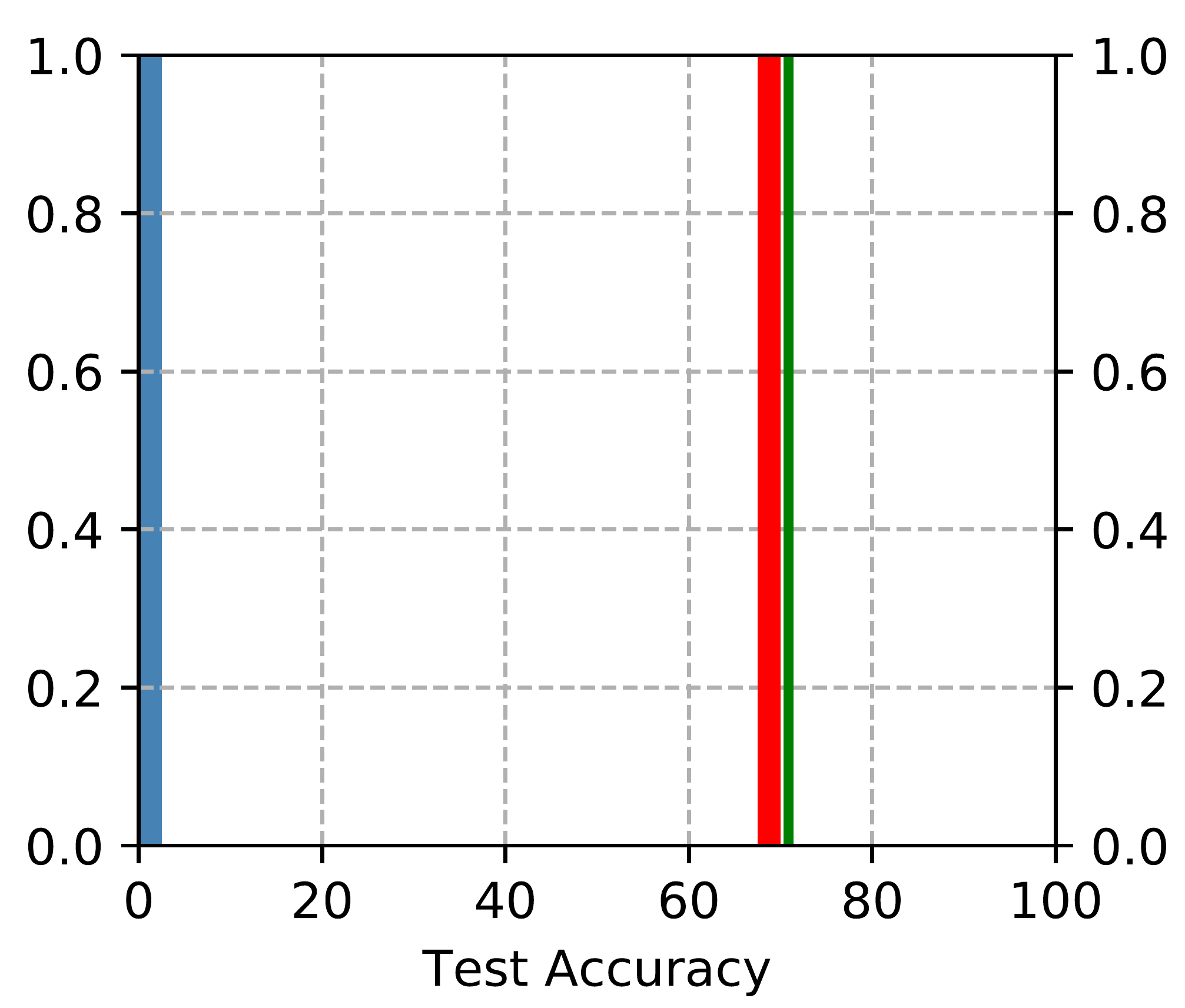}
			\caption{VGGNet}
		\end{subfigure}
		\begin{subfigure}{.3\linewidth}
			\centering
			\includegraphics[keepaspectratio=true, scale = 0.25]{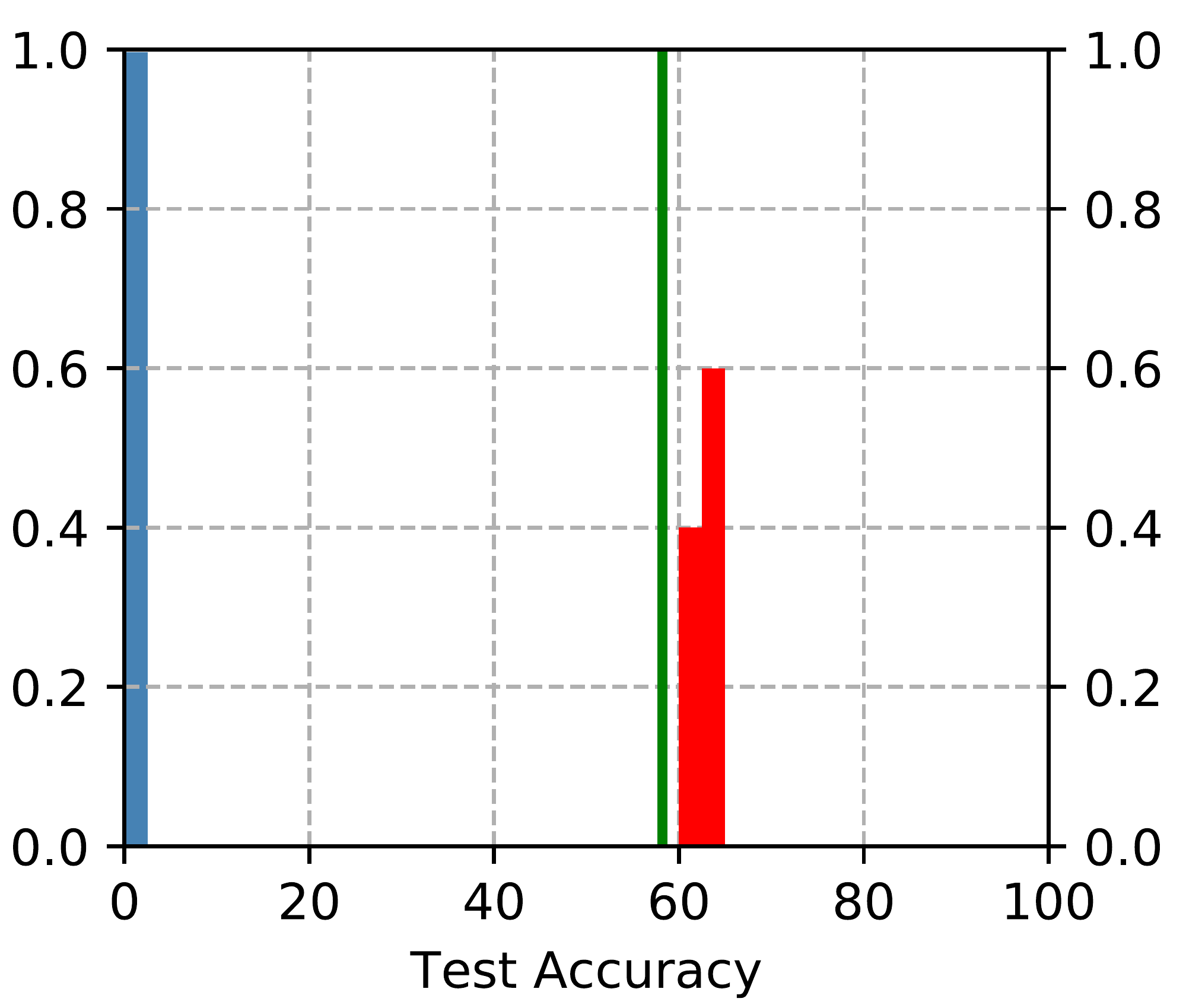}
			\caption{AlexNet}
		\end{subfigure}
		
		\caption{CIFAR100: Distributions of the test accuracies (\%) with \textit{VGGNet} and \textit{AlexNet} using Passport function V3 with random image passports.
		}
		\label{fig:hist-resnet18-alexnet-vgg16-cifar100}
	\end{figure*}

	This section describes distributions of the test accuracies with different networks on CIFAR10 (Figure \ref{fig:hist-alexnet-vgg16-cifar10}) and CIFAR100 (Figure \ref{fig:hist-resnet18-alexnet-vgg16-cifar100}), respectively. For each histogram, \textbf{green line} represents test accuracy of the original network, \textbf{red histogram} represents test accuracies with the correct passports and \textbf{blue histogram} represents test accuracies with fake passports.

\section{Discussions and Conclusions}

We renovated the paradigm in recent studies of digital watermarking for deep neural network protections, by proposing to paralyze network functionalities for unauthorized usages, and thus, preventing IP infringements in a cost-effective, proactive and timely manner. The proposed generic solution and implementation schemes are proved, by extensive experiment results, to be effective, reliable and resilient against tens of thousands of fake passport attacks and revere-engineering attacks.  

We believe this paper puts forward a new research direction for the study of deep neural networks IP protection which is urgently needed. Our  future works include the passport protection of other network architectures such as GANs, which is feasible according to the generic solution principle, yet remains to be investigated empirically.

\newpage
\balance
\bibliography{../bib/nnpass_related,../bib/NeuralNetworkCollection}

\begin{thebibliography}{15}
\providecommand{\natexlab}[1]{#1}
\providecommand{\url}[1]{\texttt{#1}}
\expandafter\ifx\csname urlstyle\endcsname\relax
  \providecommand{\doi}[1]{doi: #1}\else
  \providecommand{\doi}{doi: \begingroup \urlstyle{rm}\Url}\fi

\bibitem[Adi et~al.(2018)Adi, Baum, Cisse, Pinkas, and
  Keshet]{TurnWeakStrength_Adi2018arXiv}
Adi, Y., Baum, C., Cisse, M., Pinkas, B., and Keshet, J.
\newblock Turning your weakness into a strength: Watermarking deep neural
  networks by backdooring.
\newblock In \emph{27th USENIX Security Symposium (USENIX)}, 2018.

\bibitem[{Chen} et~al.(2018){Chen}, {Darvish Rohani}, and
  {Koushanfar}]{DeepMarks_2018arXiv}
{Chen}, H., {Darvish Rohani}, B., and {Koushanfar}, F.
\newblock {DeepMarks: A Digital Fingerprinting Framework for Deep Neural
  Networks}.
\newblock \emph{arXiv e-prints}, art. arXiv:1804.03648, April 2018.

\bibitem[{Darvish Rouhani} et~al.(2018){Darvish Rouhani}, {Chen}, and
  {Koushanfar}]{DeepSigns_2018arXiv}
{Darvish Rouhani}, B., {Chen}, H., and {Koushanfar}, F.
\newblock {DeepSigns: A Generic Watermarking Framework for IP Protection of
  Deep Learning Models}.
\newblock \emph{arXiv e-prints}, art. arXiv:1804.00750, April 2018.

\bibitem[He et~al.(2015)He, Zhang, Ren, and Sun]{he2015delving}
He, K., Zhang, X., Ren, S., and Sun, J.
\newblock Delving deep into rectifiers: Surpassing human-level performance on
  imagenet classification.
\newblock In \emph{Proceedings of the IEEE International Conference on Computer
  Vision (ICCV)}, pp.\  1026--1034, 2015.

\bibitem[He et~al.(2016)He, Zhang, Ren, and Sun]{he2016deep}
He, K., Zhang, X., Ren, S., and Sun, J.
\newblock Deep residual learning for image recognition.
\newblock In \emph{Proceedings of the IEEE Conference on Computer Vision and
  Pattern Recognition (CVPR)}, pp.\  770--778, 2016.

\bibitem[Ioffe \& Szegedy(2015)Ioffe and Szegedy]{BatchNorm_IoffeS15}
Ioffe, S. and Szegedy, C.
\newblock Batch normalization: accelerating deep network training by reducing
  internal covariate shift.
\newblock In \emph{Proceedings of the 32nd International Conference on
  International Conference on Machine Learning (ICML)}, pp.\  448--456, 2015.

\bibitem[Jia \& Potkonjak(2018)Jia and Potkonjak]{8587745}
Jia, G. and Potkonjak, M.
\newblock Watermarking deep neural networks for embedded systems.
\newblock In \emph{2018 IEEE/ACM International Conference on Computer-Aided
  Design (ICCAD)}, pp.\  1--8, 2018.

\bibitem[Krizhevsky et~al.(2012)Krizhevsky, Sutskever, and
  Hinton]{krizhevsky2012imagenet}
Krizhevsky, A., Sutskever, I., and Hinton, G.~E.
\newblock Imagenet classification with deep convolutional neural networks.
\newblock In \emph{Advances in Neural Information Processing Systems
  (NeurIPS)}, pp.\  1097--1105, 2012.

\bibitem[{Le Merrer} et~al.(2017){Le Merrer}, {Perez}, and
  {Tr{\'e}dan}]{AdStitch_2017arXiv}
{Le Merrer}, E., {Perez}, P., and {Tr{\'e}dan}, G.
\newblock {Adversarial Frontier Stitching for Remote Neural Network
  Watermarking}.
\newblock \emph{arXiv e-prints}, art. arXiv:1711.01894, November 2017.

\bibitem[Mun et~al.(2017)Mun, Nam, Jang, Kim, and Lee]{mun2017robust}
Mun, S.-M., Nam, S.-H., Jang, H.-U., Kim, D., and Lee, H.-K.
\newblock A robust blind watermarking using convolutional neural network.
\newblock \emph{arXiv preprint arXiv:1704.03248}, 2017.

\bibitem[Simonyan \& Zisserman(2014)Simonyan and Zisserman]{simonyan2014very}
Simonyan, K. and Zisserman, A.
\newblock Very deep convolutional networks for large-scale image recognition.
\newblock \emph{arXiv preprint arXiv:1409.1556}, 2014.

\bibitem[Uchida et~al.(2017)Uchida, Nagai, Sakazawa, and
  Satoh]{EmbedWMDNN_2017arXiv}
Uchida, Y., Nagai, Y., Sakazawa, S., and Satoh, S.
\newblock Embedding watermarks into deep neural networks.
\newblock In \emph{Proceedings of the 2017 ACM on International Conference on
  Multimedia Retrieval}, pp.\  269--277, 2017.

\bibitem[Vukotić et~al.(2018)Vukotić, Chappelier, and Furon]{vedranwifs}
Vukotić, V., Chappelier, V., and Furon, T.
\newblock Are deep neural networks good for blind image watermarking?
\newblock In \emph{International Workshop on Information Forensics and Security
  (WIFS)}, pp.\  1--7, 2018.

\bibitem[Zhang et~al.(2018)Zhang, Gu, Jang, Wu, Stoecklin, Huang, and
  Molloy]{ProtectIPDNN_Zhang2018}
Zhang, J., Gu, Z., Jang, J., Wu, H., Stoecklin, M.~P., Huang, H., and Molloy,
  I.
\newblock Protecting intellectual property of deep neural networks with
  watermarking.
\newblock In \emph{Proceedings of the 2018 on Asia Conference on Computer and
  Communications Security (ASIACCS)}, pp.\  159--172, 2018.

\bibitem[Zhu et~al.(2018)Zhu, Kaplan, Johnson, and Fei-Fei]{zhu2018hidden}
Zhu, J., Kaplan, R., Johnson, J., and Fei-Fei, L.
\newblock Hidden: Hiding data with deep networks.
\newblock In \emph{European Conference on Computer Vision (ECCV)}, pp.\
  682--697, 2018.

\end{thebibliography}
\bibliographystyle{icml2019}

\end{document}